\documentstyle[elsart12,osa,epsf,eqsecnum]{revtex}
\begin{document}


\voffset1.5cm


\title{Calculating Quenching Weights}
\author{Carlos A. Salgado and Urs Achim Wiedemann}

\address{Theory Division, CERN, CH-1211 Geneva 23, Switzerland}
\date{\today}
\maketitle

\begin{abstract}
We calculate the probability (``quenching weight'') that a hard parton 
radiates an {\it additional} energy fraction $\Delta E$ due to scattering 
in spatially extended QCD matter. This study is based on an exact
treatment of finite in-medium path length,
it includes the case of a dynamically expanding medium, and 
it extends to the angular dependence of the medium-induced
gluon radiation pattern. All calculations are done in the multiple soft 
scattering approximation 
(Baier-Dokshitzer-Mueller-Peign\'e-Schiff--Zakharov ``BDMPS-Z''-formalism)
and in the single hard scattering approximation ($N=1$ opacity approximation). 
By comparison, we establish a simple relation between transport 
coefficient, Debye screening mass and opacity, for which both 
approximations lead to comparable results. Together with this paper, 
a CPU-inexpensive numerical subroutine for calculating quenching weights 
is provided electronically. To illustrate its applications, we discuss 
the suppression of hadronic transverse momentum spectra in nucleus-nucleus 
collisions. 
Remarkably, the kinematic constraint resulting from finite in-medium
path length reduces significantly the 
$p_\perp$-dependence of the nuclear modification factor, thus leading to
consistency with the data measured at the Relativistic Heavy Ion 
Collider (RHIC).
\end{abstract}

\section{Introduction}
\label{sec1}

Hard partons produced in nucleus-nucleus collisions at RHIC and 
LHC propagate through highly excited matter before hadronizing
in the vacuum. The resulting medium-dependence of parton 
fragmentation is expected to affect hadronic observables. 
This is of two-fold interest. First, it provides a
novel test of the space-time evolution of the perturbative
parton shower. Second, the modification of hadronic observables
due to the spatially extended, hot and dense QCD matter
allows to characterize the properties of the transient state
produced in the collision.

Gluon emission off highly virtual hard partons is an essential
component in the standard description of parton fragmentation
in elementary processes. This effect degrades the energy of the
leading parton. Recently, it has been proposed~\cite{Baier:2001yt}
that in the presence of a spatially extended medium, the additional 
medium-induced energy degradation of the leading parton can be
described by a probability $P(\Delta E)$, the so-called quenching 
weight, which is obtained from a probabilistic 
iteration of the medium-modified elementary splitting processes 
$q \to qg$ and $g \to gg$. The main purpose of the present work
is to calculate and compare this quenching weight for different 
approximations of the medium-modified splitting process, to make 
the results for $P(\Delta E)$ available as a numerical subroutine,
and to illustrate the use of this subroutine with some applications.

We start from recent calculations~\cite{Baier:1996kr,Zakharov:1996fv,Wiedemann:2000za,Gyulassy:2000er} 
of the modification of the elementary splitting processes 
$q \to qg$ and $g \to gg$ due to multiple scattering. These results
go under the name medium-induced gluon radiation. They present
limiting cases of an unique path-integral expression given in 
eq. (\ref{2.1}) below. Technically, they collect all terms to 
leading order in nuclear enhanced modifications $O(\alpha_s A^{1/3})$,
thus accounting for the leading additional interactions of the parton 
shower with the medium.

The paper is organized as follows: In section~\ref{sec2}, 
we compare the medium-induced gluon energy distribution radiated 
off a hard parton in two limits which emphasize the role of
multiple soft and single hard medium-induced scatterings,
respectively. In section~\ref{sec3} we give results for the 
quenching weights corresponding to these limits. These quenching
weights can be calculated with the numerical subroutine accompanying 
this paper. In section~\ref{sec4}, we extend these calculations to
the case of an expanding medium, and in section~\ref{sec5}, we 
discuss the extension to radiation within a finite cone.
As application, we calculate in section~\ref{sec6} in two
different approaches the suppression of hadronic transverse
momentum spectra and we compare our results to the 
nuclear modification factor measured in Au-Au collisions at the 
Relativistic Heavy Ion Collider (RHIC).

\section{Medium-induced gluon radiation from a static medium} 
\label{sec2}
The inclusive energy distribution of gluon radiation off an in-medium 
produced parton takes the form
\cite{Wiedemann:2000za,Wiedemann:2000tf,Wiedemann:correct}
\begin{eqnarray}
  \omega\frac{dI}{d\omega}
  &=& {\alpha_s\,  C_R\over (2\pi)^2\, \omega^2}\,
    2{\rm Re} \int_{\xi_0}^{\infty}\hspace{-0.3cm} dy_l
  \int_{y_l}^{\infty} \hspace{-0.3cm} d\bar{y}_l\,
   \int d{\bf u}\,  \int_0^{\chi \omega}\, d{\bf k}_\perp\, 
  e^{-i{\bf k}_\perp\cdot{\bf u}}   \,
  e^{ -\frac{1}{2} \int_{\bar{y}_l}^{\infty} d\xi\, n(\xi)\,
    \sigma({\bf u}) }\,
  \nonumber \\
  && \times {\partial \over \partial {\bf y}}\cdot
  {\partial \over \partial {\bf u}}\,
  \int_{{\bf y}=0}^{{\bf u}={\bf r}(\bar{y}_l)}
  \hspace{-0.5cm} {\cal D}{\bf r}
   \exp\left[ i \int_{y_l}^{\bar{y}_l} \hspace{-0.2cm} d\xi
        \frac{\omega}{2} \left(\dot{\bf r}^2
          - \frac{n(\xi) \sigma\left({\bf r}\right)}{i\, \omega} \right)
                      \right]\, .
    \label{2.1}
\end{eqnarray}
Here, ${\bf k}_\perp$ denotes the transverse momentum of the emitted gluon.
The limit $k_\perp = \vert{\bf k}_\perp\vert < \chi\, \omega$
on the transverse phase space allows to
discuss gluon emission into a finite opening angle $\Theta$,
$\chi = \sin\Theta$. For the
full angular integrated quantity, $\chi = 1$. 

The two-dimensional transverse coordinates ${\bf u}$, ${\bf y}$
and ${\bf r}$ emerge in the derivation of (\ref{2.1}) as distances
between the positions of projectile components in the amplitude
and complex conjugate amplitude. The longitudinal coordinates
$y_l$, $\bar{y}_l$ integrate over the ordered longitudinal
gluon emission points in amplitude and complex conjugate amplitude,
which emerge in time-ordered perturbation theory. These
internal integration variables play no role in the following 
discussion. They are
explained in more detail in Ref. \cite{Wiedemann:2000za}.

The radiation of hard quarks or gluons differs by 
the Casimir factor $C_R = C_F$ or $C_A$, respectively. Numerical
results are for fixed coupling constant $\alpha_s = 1/3$,
except where stated otherwise. The 
properties of the medium enter eq. (\ref{2.1}) in terms of the product 
of the time-dependent density $n(\xi)$ of scattering centers times 
the strength of a single elastic scattering $\sigma({\bf r})$. 
This dipole cross section $\sigma({\bf r})$ is given 
in terms of the elastic high-energy 
cross section $\vert a({\bf q})\vert^2$ of a single scatterer,
\begin{eqnarray}
 \sigma({\bf r}) = 2 \int \frac{d{\bf q}}{(2\pi)^2}\,
                    \vert a({\bf q})\vert^2\, 
                    \left(1 - e^{i{\bf q}\cdot {\bf r}}\right)\, .
 \label{2.2}
\end{eqnarray}
In this section, we study the energy distribution (\ref{2.1}) 
for a static medium in the limiting cases of multiple soft 
and single hard momentum transfer. Then we compare these
two limiting cases of (\ref{2.1}).

\subsection{Multiple soft scattering approximation}
\label{sec2a}
For arbitrary many soft scattering centers, the projectile performs a 
Brownian motion in transverse momentum. This dynamical limiting
case can be studied in the saddle point approximation of the
path-integral (\ref{2.1}), using\cite{Zakharov:1996fv,Zakharov:1998sv}
\begin{eqnarray}
  n(\xi)\, \sigma({\bf r}) \simeq \frac{1}{2}\, \hat{q}(\xi)\, {\bf r}^2\, .
  \label{2.3}
\end{eqnarray}
Here, $\hat{q}(\xi)$ is the transport coefficient\cite{Baier:1996sk} 
which characterizes the medium-induced transverse momentum squared 
$\langle q_\perp^2\rangle_{\rm med}$ transferred to the projectile 
per unit path length $\lambda$. For a static medium, the transport
coefficient is time-independent,
\begin{equation}
  \hat{q} = \frac{\langle q_\perp^2\rangle_{\rm med}}{\lambda}\, .
  \label{2.4}
\end{equation}
In the approximation (\ref{2.3}),
the path integral in (\ref{2.1}) is equivalent 
to that of a harmonic oscillator. The corresponding analytical 
expressions are summarized in Appendix \ref{appa}. 

{\it Qualitative arguments \cite{Baier:2002tc}:}
We consider a gluon in the hard parton 
wave function. This gluon is emitted due to multiple scattering
if it picks up sufficient transverse momentum to decohere from the partonic
projectile. For this, the average phase $\varphi$ accumulated by
the gluon should be of order one,
\begin{equation}
  \varphi = \Bigg\langle \frac{k_\perp^2}{2\omega}\, \Delta z \Bigg\rangle
  \sim \frac{\hat{q}\, L}{2\omega} L = \frac{\omega_c}{\omega}\, .
  \label{2.5}
\end{equation}
Thus, for a hard parton traversing a finite path length $L$ in the medium,
the scale of the radiated energy distribution is set by 
the ``characteristic gluon frequency''
\begin{equation}
  \omega_c = \frac{1}{2}\, \hat{q}\, L^2\, .
  \label{2.6}
\end{equation}
For an estimate of the shape of the energy distribution, we 
consider the number $N_{\rm coh}$ of scattering centers which 
add coherently in the gluon phase (\ref{2.5}), 
$k_\perp^2 \simeq N_{\rm coh}\, \langle q_\perp^2\rangle_{\rm med}$. 
Based on expressions
for the coherence time of the emitted gluon, 
$t_{\rm coh} \simeq \frac{\omega}{k_\perp^2} \simeq 
\sqrt{\frac{\omega}{\hat{q}}}$
and $N_{\rm coh} = \frac{t_{\rm coh}}{\lambda} = 
\sqrt{\frac{\omega}{\langle q_\perp^2\rangle_{\rm med}\, \lambda}}$, 
one estimates for the
gluon energy spectrum per unit path length
\begin{equation}
  \omega \frac{dI}{d\omega\, dz} \simeq 
  \frac{1}{N_{\rm coh}}\, 
  \omega \frac{dI^{\rm 1\, scatt}}{d\omega\, dz} \simeq
  \frac{\alpha_s}{t_{\rm coh}}
  \simeq \alpha_s\, \sqrt{\frac{\hat{q}}{\omega}}\, .
  \label{2.7}
\end{equation}
This $1/\sqrt{\omega}$-energy dependence of the
medium-induced non-abelian gluon energy spectrum is
expected for sufficiently small $\omega < \omega_c$.

{\it Quantitative analysis:}
The gluon energy distribution (\ref{2.1}) depends not only on $\omega_c$, 
but also on the constraint $k_\perp < \chi \omega$ on the transverse 
momentum phase space of the emitted gluon. This enters the calculation
via the dimensionless kinematic constraint\cite{Salgado:2002cd}
\begin{equation}
  R_{\chi} = \frac{1}{2}\, \hat{q}\, \chi^2\, L^3\, ,
  \qquad R \equiv R_{\chi = 1} = \omega_c\, L\, .
  \label{2.8}
\end{equation}
This constraint is neglected in the argument leading to 
the $1/\sqrt{\omega}$-energy dependence of (\ref{2.7}). In the
following sections, we limit the discussion to angular fully 
integrated quantities for which $\chi = 1$. The only exception 
will be the discussion of
the angular $\Theta$-dependence of $\omega \frac{dI}{d\omega}$ in 
section \ref{sec5}, where we use $\chi = \sin\Theta$. 

The limit $R \to \infty$ which removes the kinematic constraint
from (\ref{2.1}) is either realized by extending the $k_\perp$-integration
ad hoc to infinity. Alternatively, 
$R\to \infty$ can be viewed as the limit of infinite in-medium
path length since it corresponds to $L\to \infty$ for $\chi$ and
$\omega_c$ fixed. In Appendix \ref{appa}, we derive the $R\to \infty$ 
limit of the energy distribution (\ref{2.1}),
\begin{equation}
  \lim_{R\to \infty}\, 
   \omega \frac{dI}{d\omega} =
   \frac{2\alpha_s C_R}{\pi}\, 
   \ln \Bigg \vert
   {\cos\left[\,(1+i)\sqrt{\frac{\omega_c}{2\omega}}\,\right]}
   \Bigg \vert \, .
   \label{2.9}
\end{equation}
This coincides with the result of Baier, Dokshitzer, Mueller,
Peign\'e and Schiff\cite{Baier:1996sk}.
As expected from the estimates in (\ref{2.5})
and (\ref{2.7}), it shows a characteristic $1/\sqrt{\omega}$-energy 
dependence for small $\omega$ which is suppressed above the 
characteristic gluon frequency $\omega_c$:\cite{Baier:2001yt}
\begin{eqnarray}
   \lim_{R\to \infty}\, 
   \omega \frac{dI}{d\omega} \simeq 
           \frac{2\alpha_s C_R}{\pi} 
          \left\{ \begin{array} 
                  {r@{\qquad  \hbox{for}\quad}l}                 
                  \sqrt{\frac{\omega_c}{2\, \omega}}
                  & \omega < \omega_c\, , \\ 
                  \frac{1}{12} 
                  \left(\frac{\omega_c}{\omega}\right)^2
                  & \omega > \omega_c \, .
                  \end{array} \right.
  \label{2.10}
\end{eqnarray}
The average parton energy loss is the zeroth moment of this
energy distribution, 
\begin{equation}
 \langle \Delta E \rangle_{R\to\infty} = 
 \lim_{R\to \infty}\,  \int_0^\infty d\omega\, 
   \omega \frac{dI}{d\omega}
 = \frac{\alpha_s C_R}{2}\, \omega_c\, . 
 \label{2.11}
\end{equation}
This is the well-known $L^2$-dependence of the average energy 
loss~\cite{Baier:1996kr,Baier:1996sk,Zakharov:1997uu}.
Due to the steep fall-off at large $\omega$,
the $\omega$-integral in (\ref{2.11}) is dominated by the region
$\omega < \omega_c/\sqrt{2}$. 

\begin{figure}[h]\epsfxsize=10.7cm
\centerline{\epsfbox{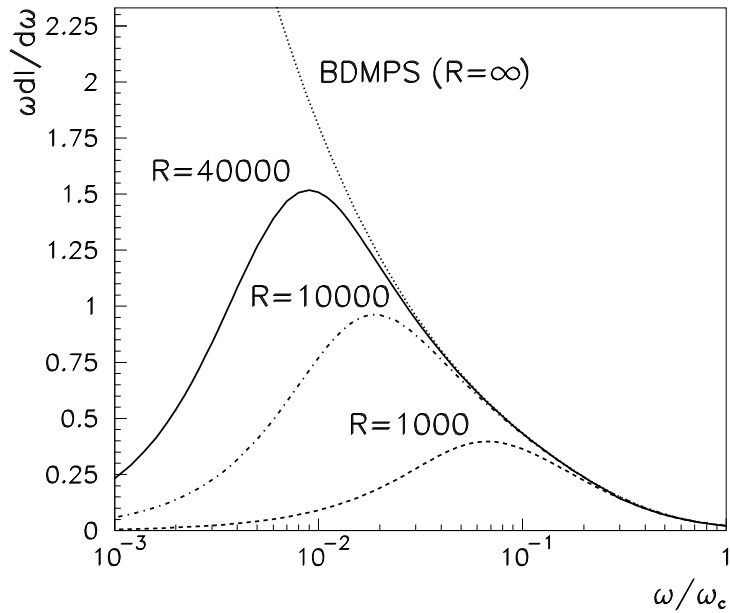}}
\vspace{0.5cm}
\caption{The medium-induced gluon energy distribution 
$\omega \frac{dI}{d\omega}$ in the multiple soft scattering
approximation for different values of the 
kinematic constraint $R = \omega_c\, L$.
}\label{fig1}
\end{figure}

\begin{figure}[h]\epsfxsize=10.7cm
\centerline{\epsfbox{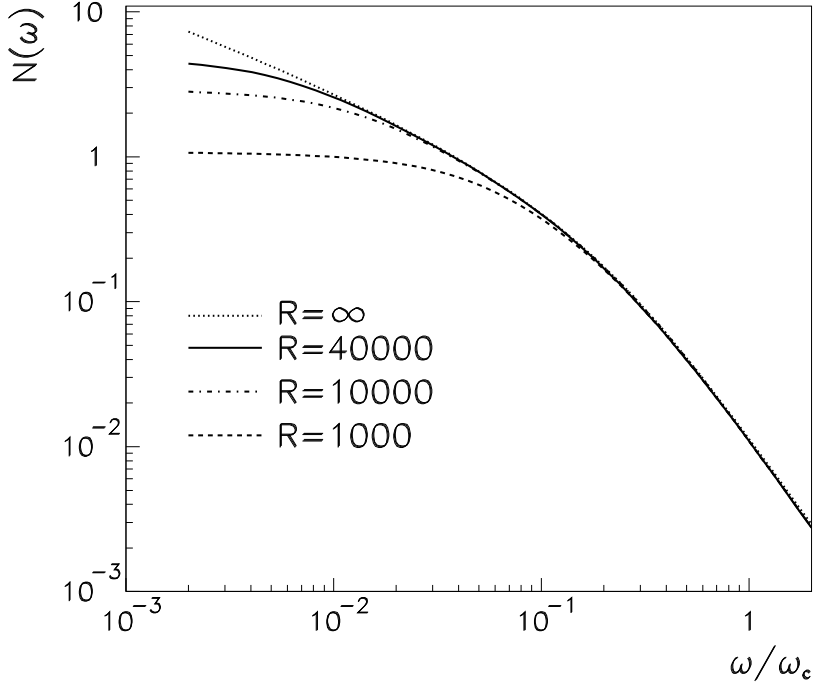}}
\vspace{0.5cm}
\caption{The multiplicity of additional medium-induced gluons 
(\protect\ref{2.14}) radiated with energy large than $\omega$.
Calculation done in the multiple soft scattering approximation.
}\label{fig2}
\end{figure}

We have evaluated numerically the energy distribution (\ref{2.1})
for finite values of the density parameter $R$. As seen in 
Fig.~\ref{fig1}, the distribution approaches for any value of $R$
the BDMPS limit (\ref{2.9}) at sufficiently large gluon energy. Below
a critical gluon energy $\hat{\omega}$, however, the finite size
gluon spectrum is depleted in comparison to the BDMPS limit.
To understand this effect, we consider the characteristic 
angle $\Theta_c$ at which medium-induced gluons are radiated on average,
\begin{equation}
  \Theta_c^2 \simeq \frac{k_\perp^2}{\omega^2}
  \simeq \frac{\sqrt{\omega \hat{q}}}{\omega^2}
  \simeq \left( \frac{\omega}{\omega_c}\right)^{-3/2}
  \frac{1}{R}\, .
  \label{2.12}
\end{equation}
For $\Theta_c \sim 1$, the emitted gluons are sensitive to the 
kinematic constraint since $k_\perp \sim O(\omega)$. The condition 
$\Theta_c \sim 1$ thus provides an estimate for 
the gluon energy $\hat{\omega}$ below which the energy distribution
is cut off,
\begin{equation}
  \frac{\hat{\omega}}{\omega_c} \propto 
        \left(\frac{1}{R} \right)^{2/3}\, .
  \label{2.13}
\end{equation}
The position of the maximum of $\omega \frac{dI}{d\omega}$ 
as a function of $R$ is consistent with this dependence
on $\hat{\omega}$, see
Fig.~\ref{fig1}. In general, gluon radiation at small energies 
corresponds to gluon radiation at large angle
and is depleted as soon as the finite size of the transverse 
momentum phase space becomes relevant. 
This suppression of the non-perturbative small-$\omega$ contributions 
helps to make the calculation of medium-induced energy loss 
perturbatively stable.
\vspace{-0.5cm}
%
The gluon energy distribution (\ref{2.1}) also determines
the multiplicity $N(\omega)$ of gluons emitted with energies 
larger than $\omega$
\begin{equation}
  N(\omega) \equiv \int_\omega^\infty d\omega'\,
                    \frac{dI(\omega')}{d\omega'}\, .
  \label{2.14}
\end{equation}
In the absence of kinematic constraints, and for sufficiently
small energies $\omega$, one finds from (\ref{2.10}) that the 
total multiplicity diverges like $1/\sqrt{\omega}$,\cite{Baier:2001yt}
\begin{equation}
  \lim_{R\to\infty} N(\omega)
             = \frac{2 \alpha_s\, C_R}{\pi} 
               \sqrt{\frac{2\omega_c}{\omega}}\, , 
               \qquad \hbox{for}\, \quad \omega < \omega_c\, .
  \label{2.15}
\end{equation}
However, realistic kinematic constraints on the transverse
momentum phase space ($R < \infty$) deplete the gluon energy
distribution at small $\omega$ and ensure that the
total gluon multiplicity $N(\omega=0)$ is finite, see Fig.~\ref{fig2}.

For realistic kinematic constraints, $R < 10000$, the average
additional total multiplicity is $N(\omega = 0) \leq 3$. In comparison
to the typically $\sim 5- 10$ semi hard partons which are the partonic
final state of a 100 GeV jet simulated in a parton shower, this 
additional multiplicity is not negligible.   
It supports the naive expectation that the number of
partons in the jet increases and softens with increasing transport
coefficient or path length.

%

\subsection{Single hard scattering approximation}
\label{sec2b}
In the previous subsection, we have studied the energy distribution 
(\ref{2.1}) of medium-induced gluon radiation in the limit in which 
the partonic projectile performs a transverse Brownian motion due to
multiple soft scattering. Now, we consider the opposite limiting
case in which the radiation pattern results from an incoherent
superposition of very few $n_0L$ single hard scattering processes
positioned within path length $L$. This limit is obtained
by expanding the integrand of the energy distribution (\ref{2.1}) in 
powers of $\left( n(\xi)\, \sigma({\bf r})\right)^N$ up
to first order\cite{Wiedemann:2000za,Gyulassy:2000er,Gyulassy:2000fs}. 
Analytical expressions are given in Appendix \ref{appb}.

{\it Qualitative arguments:} We consider a hard partonic projectile
which picks up a single transverse momentum $\mu$ by interacting with a 
single hard scatterer. An additional gluon of energy $\omega$ decoheres
from the projectile wave function if its typical formation
time $\bar{t}_{\rm coh} = \frac{2\omega}{\mu^2}$ is smaller than the
typical distance $L$ between the production point of the parton
and the position of the scatterer. The relevant phase is 
\begin{equation}
  \gamma = \frac{L}{\bar{t}_{\rm coh}} \equiv \frac{\bar{\omega}_c}{\omega}
  \, ,
  \label{2.16}
\end{equation}
which indicates a suppression of gluons with energy $\omega$
larger than the characteristic gluon energy 
\begin{equation}
  \bar\omega_c = \frac{1}{2} \mu^2\, L\, .
  \label{2.17}
\end{equation}
The gluon energy spectrum per unit path length can be estimated
in terms of the coherence time $\bar{t}_{\rm coh}$ and of the
average number $n_0\, L$ of scattering centers contributing
incoherently
\begin{equation}
  \omega \frac{dI^{N=1}}{d\omega\, dz} \simeq 
  (n_0\, L)\, \frac{\alpha_s}{\bar{t}_{\rm coh}}
  \simeq (n_0\, L)\, \alpha_s\, \frac{\mu^2}{\omega}\, .
  \label{2.18}
\end{equation}
This is the typical $1/\omega$-dependence of the non-abelian
gluon radiation spectrum in the absence of LPM-type
destructive interference effects.

{\it Quantitative analysis:} We have calculated the first order
in opacity $n_0L$ of the gluon energy distribution (\ref{2.1}). To
first order, the entire medium-dependence comes from the interaction 
of the hard parton with a single static scattering center, multiplied
by the number $n_0L = L/\lambda$ of scattering centers along the
path. Modeling the single scatterer by a Yukawa potential with 
Debye screening mass $\mu$,
we derive in Appendix \ref{appb}
\begin{eqnarray}
  \omega \frac{dI^{N=1}}{d\omega} &=& 2 \frac{\alpha_s\, C_R}{\pi}\,
   (n_0L)\, \gamma\,   
  \int_0^\infty  dr\,  \frac{r - sin(r)}{r^2}
                 \nonumber \\   
  && \times
  \left( \frac{1}{r + \gamma} - 
         \frac{1}{\sqrt{( (\bar{R}/2\gamma) + r + \gamma)^2
                       - 4 r\bar{R}/2\gamma}}\right)\, .
  \label{2.19}
\end{eqnarray}
This energy distribution depends on the phase factor $\gamma$ defined 
in (\ref{2.16}), and on the kinematic
constraint in transverse momentum phase space,
\begin{equation}
  \bar{R}_{\chi} = \frac{1}{2}\, \chi^2 \mu^2\, L^2\, ,\qquad
  \bar{R} \equiv \bar{R}_{\chi=1} = \bar\omega_c\, L\, .
  \label{2.20}
\end{equation}
In what follows, we work for $\chi=1$ except for the discussion
of the angular dependence in section \ref{sec5}.
In the limit in which the kinematic constraint is removed, the
characteristic $1/\omega$-energy dependence of the estimate (\ref{2.18})
is recovered for sufficiently large gluon energies 
$\omega > \bar\omega_c$,
\begin{eqnarray}
   \lim_{\bar{R}\to \infty}\, 
   \omega \frac{dI^{N=1}}{d\omega} &=& 
   2\, \frac{\alpha_s\, C_R}{\pi}\, \left( n_0\, L\right)\,
   \gamma\,   
  \int_0^\infty  dr\, \frac{1}{r + \gamma}\,  
                  \frac{r - sin(r)}{r^2}\, 
   \nonumber \\
   &\simeq & 
   2\, \frac{\alpha_s\, C_R}{\pi}\, \left( n_0\, L\right)\,
          \left\{ \begin{array} 
                  {r@{\qquad  \hbox{for}\quad}l}
                  \log \left[ \frac{\bar\omega_c}{\omega}\right]
                  & \bar\omega_c > \omega\, ,\\ 
                  \frac{\pi}{4}\, \frac{\bar\omega_c}{\omega}
                  & \bar\omega_c < \omega\, .  
                  \end{array} \right.
  \label{2.21}
\end{eqnarray}
This limit agrees with the results of Gyulassy, Levai and 
Vitev\cite{Gyulassy:2000fs}.
The average parton energy loss for a single hard scattering is
dominated by contributions from the region 
$\omega > \bar\omega_c$,\cite{Gyulassy:2000fs,Zakharov:2000iz}
\begin{equation}
 \lim_{\bar{R} \to \infty} \langle \Delta E \rangle^{N=1} = 
 \lim_{\bar{R} \to \infty}\,  \int d\omega\, 
   \omega \frac{dI^{N=1}}{d\omega}
   \simeq \frac{\alpha_s C_R}{2} (n_0L)\, \bar\omega_c\, 
   \log\left[ E/\bar\omega_c\right]\, . 
 \label{2.22}
\end{equation}
It is logarithmically enhanced in comparison
to the region $\omega < \bar{\omega}_c$ for which
\begin{equation}
 \lim_{\bar{R} \to \infty}\,  \int_0^{\bar{\omega}_c} d\omega\, 
   \omega \frac{dI^{N=1}}{d\omega}
   \simeq \frac{2\, \alpha_s C_R}{\pi} (n_0L)\, \bar\omega_c\, . 
 \label{2.23}
\end{equation}
%
%
\begin{figure}[h]\epsfxsize=10.7cm
\centerline{\epsfbox{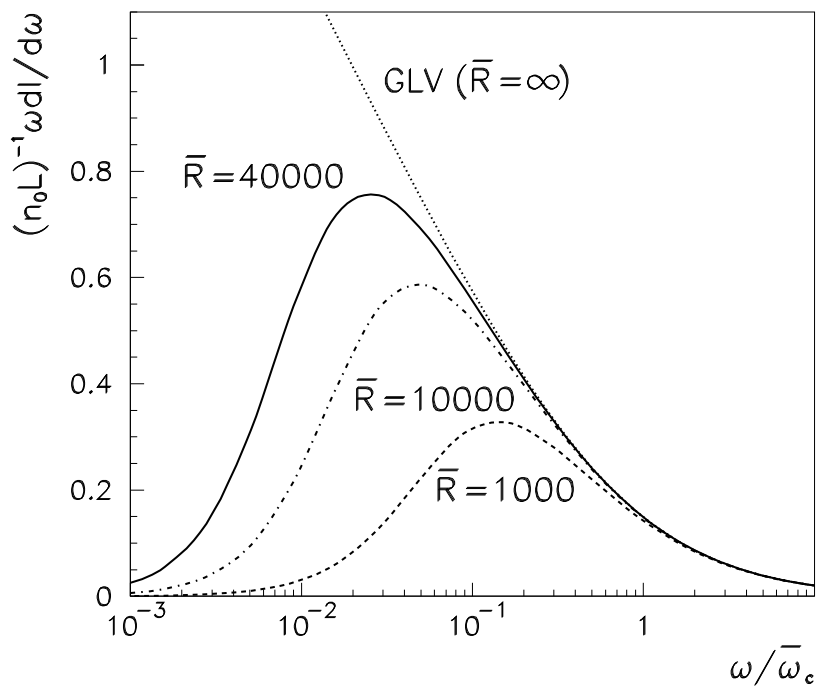}}
\vspace{0.5cm}
\caption{
The medium-induced gluon energy distribution 
$\omega \frac{dI}{d\omega}$ for a hard quark in the 
single hard scattering approximation, calculated for different
values of the kinematic constraint $\bar{R}$. 
}\label{fig3}
\end{figure}
%
Remarkably, the average parton energy
loss receives its dominant contribution from the hard region
$\omega > \bar{\omega}_c$ in the opacity approximation (\ref{2.22}) 
but from the soft region $\omega < \omega_c$ in the multiple soft
scattering approximation (\ref{2.11}).  

We have evaluated numerically the energy distribution (\ref{2.19})
for finite values of the kinematic constraint $\bar{R}$. 
In close analogy to the multiple soft scattering approximation, 
the emission of soft gluons is suppressed in the opacity approximation 
due to the kinematic constraint $\bar{R} = \bar{\omega}_c L$ on the
transverse momentum phase space, see Fig.~\ref{fig3}.
To estimate the scale $\hat{\omega}$ at which this suppression
sets in, we parallel the argument leading to eq. (\ref{2.12}). 
We require that the characteristic angle of the
gluon emission is of order one, finding
\begin{equation}
  \Theta_c^2 \simeq \frac{\mu^2}{\hat{\omega}^2}
  \simeq \left(\frac{\bar\omega_c}{\hat{\omega}} \right)^2 
         \frac{1}{\bar{R}} \sim 1\quad \Longrightarrow
         \quad \frac{\hat{\omega}}{\bar\omega_c} 
               \propto \frac{1}{\sqrt{\bar{R}}}  \, .
  \label{2.24}
\end{equation}
The numerical position of the maximum of $\omega \frac{dI^{N=1}}{d\omega}$
in Fig.~\ref{fig3} changes $ \propto \frac{1}{\sqrt{\bar{R}}}$, in
accordance with this estimate. We thus have a semi-quantitative 
understanding of how phase space constraints deplete the non-perturbative
soft region of the medium-induced gluon energy distribution.
%
\begin{figure}[h]\epsfxsize=10.7cm
\centerline{\epsfbox{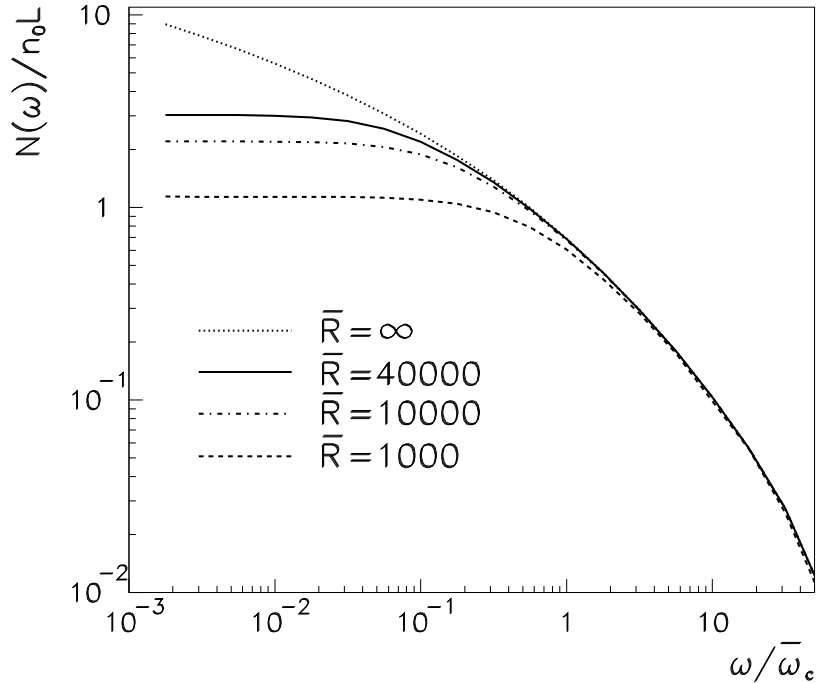}}
\vspace{0.5cm}
\caption{The multiplicity of additional medium-induced gluons 
(\protect\ref{2.14}) radiated with energy larger than $\omega$. 
Calculation done in the single hard scattering approximation.
}\label{fig4}
\end{figure}
%
In Fig.~\ref{fig4}, the additional medium-induced gluon multiplicity
(\ref{2.14}) is calculated in the opacity approximation.
In the absence of kinematic constraints $\bar{R} = \infty$ 
($R = \infty$) and for sufficiently small gluon energies 
$\omega < \bar{\omega}_c$ ($\omega < \omega_c$), this multiplicity
changes $\propto \frac{1}{\omega}$ in the opacity approximation 
($\propto \frac{1}{\sqrt{\omega}}$ in the multiple soft scattering 
approximation). In the presence of kinematic constraints, 
the total additional multiplicity is comparable for
both approximations: $N(\omega = 0) \leq 3$. 

\subsection{Comparison: multiple soft vs. single hard scattering 
approximation}
\label{sec2c}

\begin{figure}[h]\epsfxsize=10.7cm
\centerline{\epsfbox{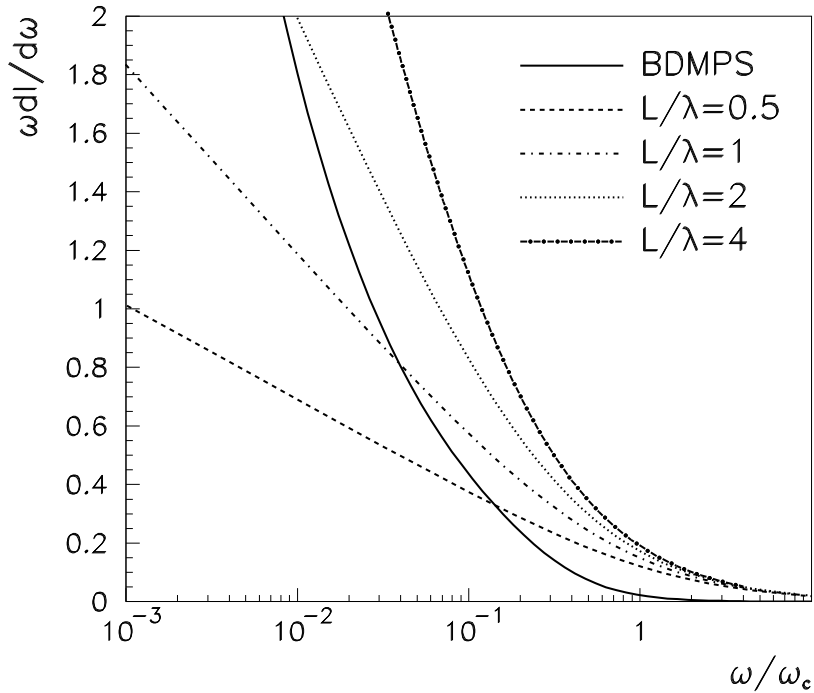}}
\vspace{0.5cm}
\caption{The gluon energy distribution without kinematic constraint
($R$, $\bar{R} \to \infty$) as calculated in the 
multiple soft scattering approximation, and in the single hard
scattering approximation for $n_0L = 0.5, 1, 2, 4$. Results for
the single hard scattering approximation 
are plotted for $(L/\lambda)\, \bar{\omega}_c = \omega_c$.
}\label{fig5}
\end{figure}

{\it Qualitative:}
The transverse momentum accumulated by a projectile due to
Brownian motion increases linearly $\propto \mu^2\, n_0\, L$ 
with path length where $n_0 = \frac{1}{\lambda}$ denotes the
longitudinal density of scattering centers. This leads to 
\begin{equation}
  \mu^2\, n_0\, L = \hat{q}\, L\, ,
  \qquad \hbox{for Brownian motion,}
  \label{2.25}
\end{equation}
and thus
\begin{equation}
  \omega_c = \frac{1}{2} \hat{q} L^2 = \frac{L}{\lambda}\, 
             \bar{\omega}_c\, ,
               \qquad \hbox{in the multiple soft scattering limit.}
  \label{2.26}
\end{equation}
Recent applications of the opacity approximation use
$1\leq \frac{L}{\lambda} \leq 3$. In this case, the gluon 
energy distribution is much harder in the opacity approximation
than in the multiple soft scattering approximation, see Fig.~\ref{fig5}. 

{\it Quantitative:} 
The relation $\omega_c = (n_0L)\, \bar{\omega}_c$ holds only if
the projectile accumulates transverse momentum by Brownian motion.
In general, deviations from Brownian motion are due to the high 
transverse momentum tails of the elastic scattering cross sections,
\begin{equation}
  \vert a({\bf q})\vert^2 = 
  \frac{ \mu^2 }{\pi ({\bf q}^2 + \mu^2)^2}\, .
  \label{2.27}
\end{equation}
In QED, the Coulomb scattering distribution is well represented 
by the theory of Moli\`ere and shows logarithmic deviations from
Brownian motion. For QCD, one can identify an analogous logarithmic 
term in the transport coefficient (\ref{2.3}) by
expanding the dipole cross section (\ref{2.2}),
\begin{equation}
  \hat{q}\, L = n_0 L \int \frac{d^2{\bf q}}{(2\pi)^2}\, 
  \vert a({\bf q})\vert^2\, \frac{1}{2}\, {\bf q}^2\, 
  \cos^2\varphi \sim (n_0 L)\, \mu^2\, \ln
  \sqrt{\frac{E_{\rm cut}}{\mu}}\, .
  \label{2.28}
\end{equation}
Here, $E_{\rm cut}$ denotes the upper cut-off of the logarithmically
divergent $q$-integral. This changes eq.~(\ref{2.26}) to 
\begin{equation}
  \omega_c = (n_0L)\, \bar{\omega_c} \ln \sqrt{\frac{E_{\rm cut}}{\mu}}\, .
  \label{2.29}
\end{equation}
The logarithmic term makes the comparison between single hard
and multiple soft scattering approximation more difficult. Based on
(\ref{2.29}), the curves for the single hard scattering approximation
should be shifted in Fig.~\ref{fig5} by 
a factor $\ln \sqrt{\frac{E_{\rm cut}}{\mu}} > 1$ to the left. For
realistic values [$\mu \geq \Lambda_{\rm QCD}$ and
$E_{\rm cut} \leq E$ say], we find 
$\ln \sqrt{\frac{E_{\rm cut}}{\mu}} \ll 10$. Thus,
the above conclusion stays unchanged: the medium-induced gluon 
energy distribution is significantly harder in the single hard
scattering approximation than in the multiple soft one.
%
%
\section{Quenching Weights}
\label{sec3}

Medium-induced gluon radiation modifies the correspondence
between the initial parton and the final hadron momenta. 
This modification can be determined from the distribution $P(\Delta E)$ of 
the additional medium-induced energy loss which we calculate now.
If gluons are emitted independently, $P(\Delta E)$
is the normalized sum of the emission probabilities
for an arbitrary number of $n$ gluons which carry away the total
energy $\Delta E$:
\begin{eqnarray}
  P(\Delta E) = \sum_{n=0}^\infty \frac{1}{n!}
  \left[ \prod_{i=1}^n \int d\omega_i \frac{dI(\omega_i)}{d\omega}
    \right]
    \delta\left(\Delta E - \sum_{i=1}^n \omega_i\right)
    \exp\left[ - \int_0^\infty \hspace{-0.2cm}
      d\omega \frac{dI}{d\omega}\right]\, .
   \label{3.1}
\end{eqnarray}
The summation over arbitrarily many gluon emissions in 
(\ref{3.1}) can be performed by Laplace transformation~\cite{Baier:2001yt}
\begin{eqnarray}
  P(\Delta E) &=& \int_C \frac{d\nu}{2\pi i}\, {\cal P}(\nu)\,
  e^{\nu\Delta E}\, ,
  \label{3.2}\\
  {\cal P}(\nu) &=& \exp\left[ -\int_0^\infty
    d\omega\, \frac{dI(\omega)}{d\omega}\,
    \left(1- e^{-\nu\, \omega}\right)\right]\, .
  \label{3.3}
\end{eqnarray}
Here, the contour $C$ runs along the imaginary axis with
${\rm Re}\nu = 0$. In general, the probability distribution
$P(\Delta E)$ has a discrete and a continuous 
part,\cite{Salgado:2002cd}
\begin{equation}
  P(\Delta E) = p_0\, \delta(\Delta E) + p(\Delta E)\, .
   \label{3.4}
\end{equation}
The discrete weight $p_0$ may be viewed as the probability that 
no additional gluon is emitted due to in-medium scattering 
and hence no medium-induced energy loss occurs. This weight
is determined by the total gluon multiplicity,
\begin{equation}
  p_0 = \lim_{\nu \to \infty}\, {\cal P}(\nu) = 
  \exp \left[ -N(\omega = 0)\right]\, .
  \label{3.5} 
\end{equation}  
For finite in-medium path length, there is always a finite
probability $p_0\not= 0$ that the projectile is not affected by
the medium. Only a finite number of additional medium-induced 
gluons can be emitted, see (\ref{3.5}). For infinite in-medium
path length, one finds 
\begin{equation}
 \lim_{R\to\infty} p_0 = 0\, .
  \label{3.6}
\end{equation}
The medium-induced gluon energy distribution $\omega \frac{dI}{d\omega}$ 
determines to what extent the total energy distribution of a given
parton deviates from its ``vacuum'' fragmentation in an elementary 
collision,
\begin{equation}
 \omega \frac{dI^{\rm (tot)}}{d\omega} = 
 \omega \frac{dI^{\rm (vac)}}{d\omega} +
 \omega \frac{dI}{d\omega}\, .
 \label{3.7}
\end{equation}
From the Laplace transform (\ref{3.2}), we obtain for the corresponding
total probability
\begin{equation}
  P^{\rm (tot)}(\Delta E) = \int_0^\infty  d\bar{E} \, 
  P(\Delta E - \bar{E}) \, 
   P^{\rm (vac)}(\bar{E})\, . 
   \label{3.8}
\end{equation} 
The probability $P^{\rm (tot)}(\Delta E)$ is normalized to unity and it is 
positive definite. In contrast, the medium-induced modification of this 
probability, $P(\Delta E)$, is a generalized probability. It can take
negative values for some range in $\Delta E$, as long as its normalization
is unity,
\begin{equation}
  \int_0^\infty  d\bar{E} \,  P(\bar{E})
  = p_0 + \int_0^\infty  d\bar{E} \,  p(\bar{E}) = 1\, .
  \label{3.9}
\end{equation}
In this section, we calculate $P(\Delta E)$ in the multiple soft
and single hard scattering approximations. The results of these
quenching weights are available as a FORTRAN routine \cite{carlosweb}. 

\subsection{Quenching weights in the multiple soft scattering approximation}
\label{sec3a}
By numerical evaluation of the Laplace transform (\ref{3.2}), (\ref{3.3}),
we have calculated quenching weights $P(\Delta E)$ for the 
medium-induced energy distribution (\ref{2.1}) in the multiple
soft scattering approximation. To motivate the range of parameter
values studied in what follows, we relate the transport coefficient 
$\hat{q}$ to the in-medium path length $L$ and the saturation 
scale $Q_s$,\cite{Baier:2002tc}
\begin{equation}
  Q_s^2 \simeq \hat{q}\, L\, .
  \label{3.10}
\end{equation}
\vspace{-1.5cm}
\begin{figure}[h]\epsfxsize=10.7cm
\centerline{\epsfbox{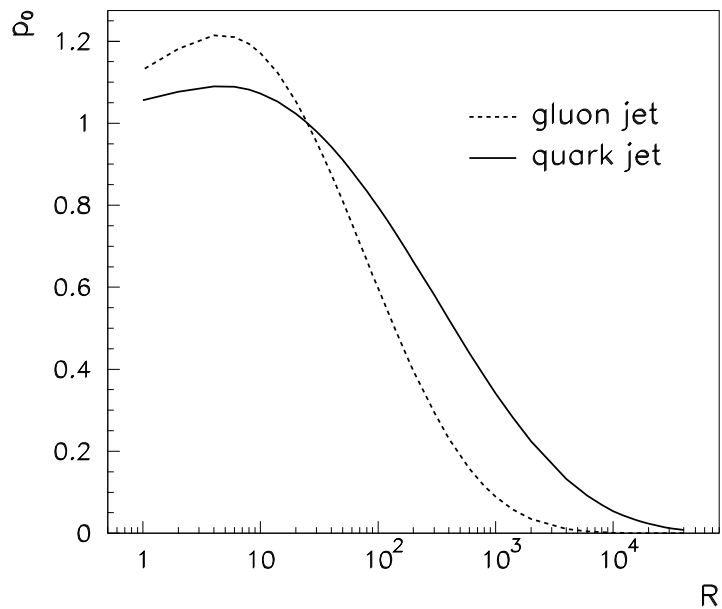}}
\vspace{0.5cm}
\caption{The discrete part $p_0$ of the quenching weight 
(\protect\ref{3.4}) calculated in the multiple soft scattering 
approximation as a function of $R=\omega_c\, L$. 
}\label{fig6}
\end{figure}
%
Estimates for the saturation scale are very uncertain but 
$Q_s^2 \leq (3\, {\rm GeV})^2$ may be considered as an upper bound at LHC.
To discuss in-medium path lengths $L$ up to twice a nuclear Pb
radius, we thus have to explore the parameter space up to $R < 40000$.
We choose a very small lower value, $R = \chi^2 \omega_c\, L^3 = 1$, 
in order to tabulate quenching weights for the radiation outside
very small opening angles $\chi = \sin\Theta$. 
All results will be given for energies in units of $\omega_c$.
\vspace{-1.0cm}
\begin{figure}[h]\epsfxsize=14.0cm
\centerline{\epsfbox{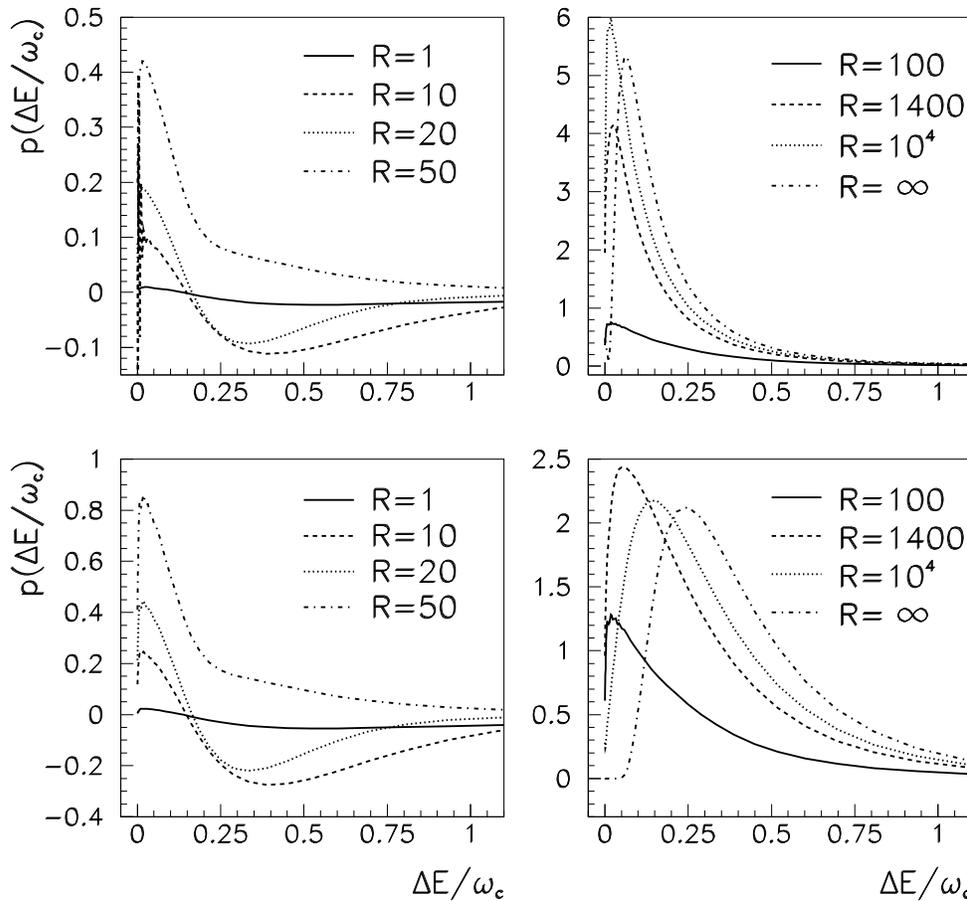}}
\caption{The continuous part of the quenching weight 
(\protect\ref{3.4}), calculated in the multiple soft scattering 
approximation for a hard quark (upper row) or hard gluon (lower row).
}\label{fig7}
\end{figure}
%
The discrete weight $p_0$ of the probability distribution 
$P(\Delta E)$ is plotted in Fig.~\ref{fig6} as a function of the 
kinematic cut-off $R = \omega_c\, L$. It approaches unity in
the absence of a medium ($R\to 0$) and it vanishes in the
limit of infinite in-medium path length, see (\ref{3.6}).
Remarkably, $p_0$ exceeds unity for small values $R < 100$. 
This indicates that there is a phase space region at very small 
transverse momentum, into which {\it less} gluons 
are emitted in the medium than in the vacuum. The ``vacuum'' 
gluon radiation is shifted to larger transverse momentum in the
presence of a medium\cite{Wiedemann:2000tf}.  The decrease of the 
discrete weight for large $R$ and its growth above unity 
for sufficiently small $R$ both depend on the strength of 
the interaction between partonic projectile and medium. 
They are thus more pronounced for gluons than for quarks. 

The continuous part $p(\Delta E)$ of the probability distribution 
(\ref{3.4}) is shown in  Fig.~\ref{fig7} as a function of the
dimensionless energy fraction $\Delta E/\omega_c$ for different 
values of the kinematic constraint $R$. Increasing the density 
of the medium (i.e. increasing the transport coefficient $\hat{q}$) 
or increasing the in-medium path length $L$ corresponds to an increase of 
$\omega_c$ and $R$. The Figs. \ref{fig6} and \ref{fig7} specify
how the probability that the parton loses an energy fraction 
$\Delta E$ changes with these medium properties. As expected
from the normalization (\ref{3.9}), the continuous part $p(\Delta E)$ 
shows predominantly negative contributions for small values $R < 100$
where the discrete weight $p_0$ exceeds unity. 

In the limit $R\to \infty$, the quenching weight was found to be
fit very well by a two-parameter log-normal distribution\cite{Arleo:2002kh}. 
This is a heuristic observation which is difficult to connect to
the analytical structure of the gluon energy distribution.
Analytically, an estimate of the quenching 
weight can be obtained\cite{Baier:2001yt} in the limit $R\to \infty$ 
from the small-$\omega$ approximation 
$\omega \frac{dI}{d\omega} \propto \frac{1}{\sqrt{\omega}}$ in 
eq. (\ref{2.10}),
\begin{equation}
  P^{\rm approx}_{\rm BDMS}(\epsilon)
  = \sqrt{\frac{a}{\epsilon^3}} 
     \exp\left[-\frac{\pi\, a}{\epsilon} \right]\, ,
     \qquad \hbox{\rm where}\, \, 
     a=\frac{2\, \alpha_s^2\, C_R^2}{\pi^2}\omega_c\, .
  \label{3.11}
\end{equation}
%
This approximation is known to capture \cite{Salgado:2002cd} the 
rough shape of the probability distribution for large system size, but 
it has an unphysical large $\epsilon$-tail with infinite first moment 
$\int d\epsilon\, \epsilon\, P^{\rm approx}_{\rm BDMS}(\epsilon)$.
Also, its maximum $\epsilon_{\rm max} = \frac{2a\pi}{3}$ grows
stronger with the effective coupling $\alpha_s\, C_R$ than the 
numerical result in Fig. \ref{fig7}.
%
%
\vspace{-1.0cm}
\begin{figure}[h]\epsfxsize=10.7cm
\centerline{\epsfbox{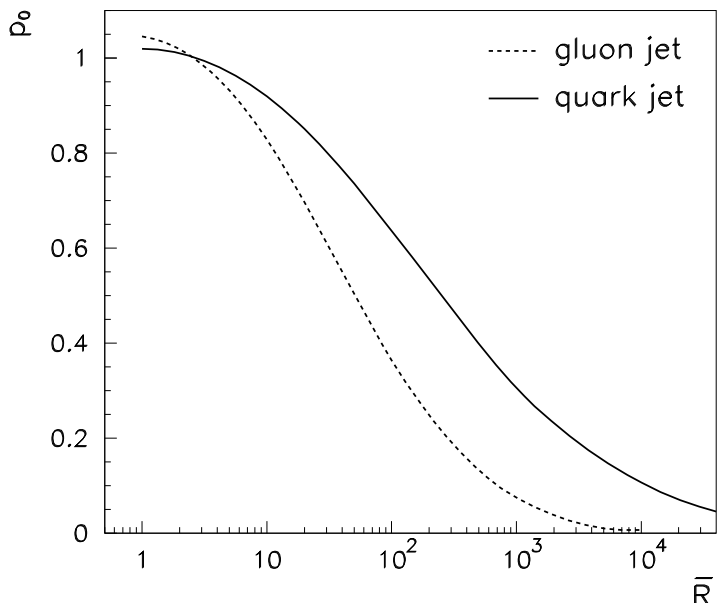}}
\caption{The discrete part $p_0$ of the quenching weight calculated
in the single hard scattering approximation for opacity $n_0L =1$. 
}\label{fig8}
\end{figure}
\subsection{Quenching Weights in the opacity approximation}
\label{sec3b}
We have evaluated the quenching weight (\ref{3.4}) for the medium-induced
gluon energy distribution in the $N=1$ opacity approximation (\ref{2.19}).
In general, the quenching weight depends in this approximation on the 
characteristic gluon energy $\bar{\omega}_c$, the kinematic 
constraint $\bar{R} = \chi^2 \bar{\omega}_c L$, and the opacity 
$n_0L$. 
\vspace{-1.0cm}
\begin{figure}[h]\epsfxsize=14.0cm
\centerline{\epsfbox{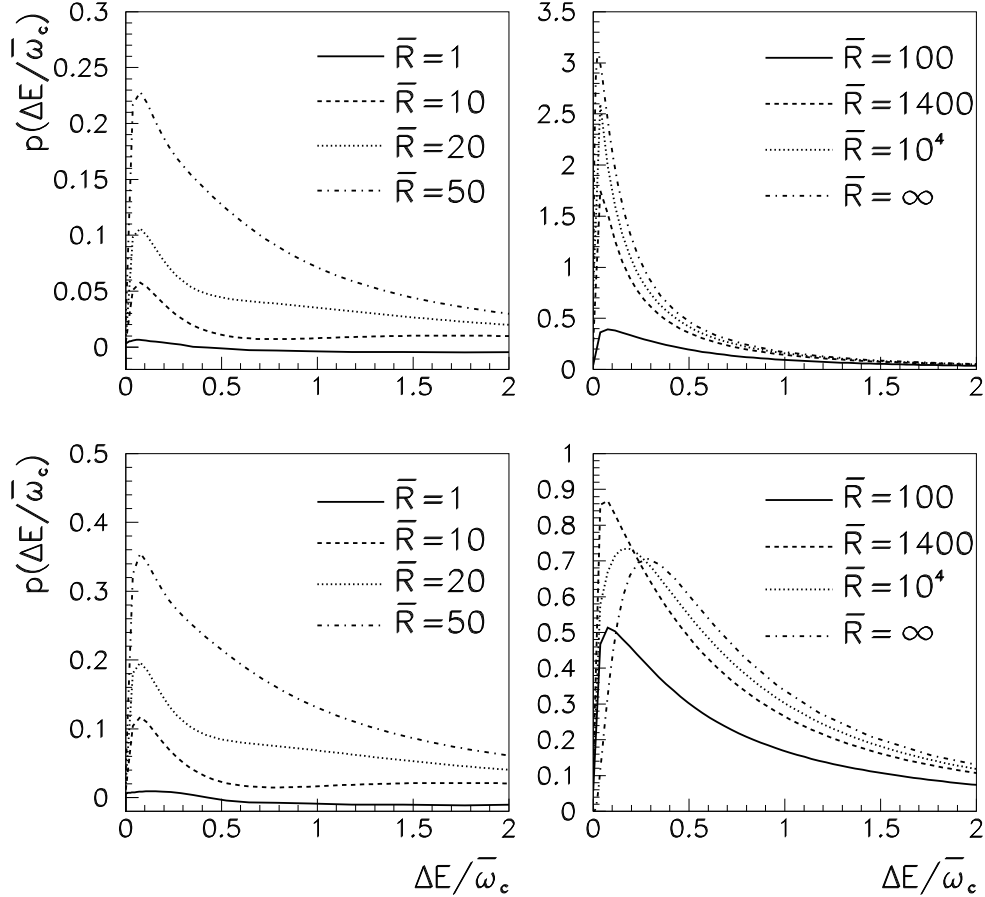}}
\caption{The continuous part of the quenching weight calculated in
the single hard scattering approximation with opacity $n_0\, L =1$ 
for a hard quark (upper row) or hard gluon (lower row).
}\label{fig9}
\end{figure}
%
For the numerical results presented in Figs. \ref{fig8} and \ref{fig9}, 
we use $n_0L =1$. The gluon energy distribution $\omega \frac{dI}{d\omega}$
depends linearly on $n_0L$, but the quenching weight shows a complicated
dependence on $n_0L$; it has to be calculated 
separately for each value of $n_0L$ from eqs. (\ref{3.2}) and (\ref{3.3}).
However, since $n_0L$ multiplies the Casimir factor $C_R$ in 
the gluon energy distribution, the quenching weight for gluons
with $n_0L =1$ is identical
to the quenching weight for quarks with $n_0L = C_A/C_F = 2.25$.
Vice versa, the quenching weight for quarks given in 
Figs. \ref{fig8} and \ref{fig9}
can be viewed as a quenching weight for gluons with 
$n_0L = C_F/C_A$.
%
\subsection{Comparison: multiple soft vs. single hard scattering approximation}
\label{sec3c}
In the opacity approximation, one specifies both the average transverse
momentum squared $\sim \mu^2\, n_0\, L$ transferred to the projectile
and the average number $n_0\, L$ of scattering centers involved in
this momentum transfer. This is in contrast to the multiple soft
scattering approximation which specifies the average transverse
momentum squared transferred to the projectile irrespective of the
number of scattering centers involved. Thus, the single hard 
scattering approximation contains one additional model parameter,
the opacity $n_0\, L$.

 Despite this difference, we want to compare the quenching weights
obtained in both approximations. To this end, we start from the
relations 
\begin{equation}
  R \simeq (n_0L)\, \bar{R}\, ,\qquad 
  \omega_c \simeq (n_0L)\, \bar{\omega}_c\, ,
  \label{3.12}
\end{equation}
discussed in section~\ref{sec2c}. Keeping the values of $R$, $\omega_c$
and $\bar{R}$, $\bar{\omega}_c$ fixed, we ``fit'' the opacity $n_0L$
such that the quenching weights obtained in both approximations show
the best agreement. This allows to discuss for both approximations 
differences in functional shape which cannot be removed by a change
of model parameters.
\vspace{-1cm}
\begin{figure}[h]\epsfxsize=11.7cm
\centerline{\epsfbox{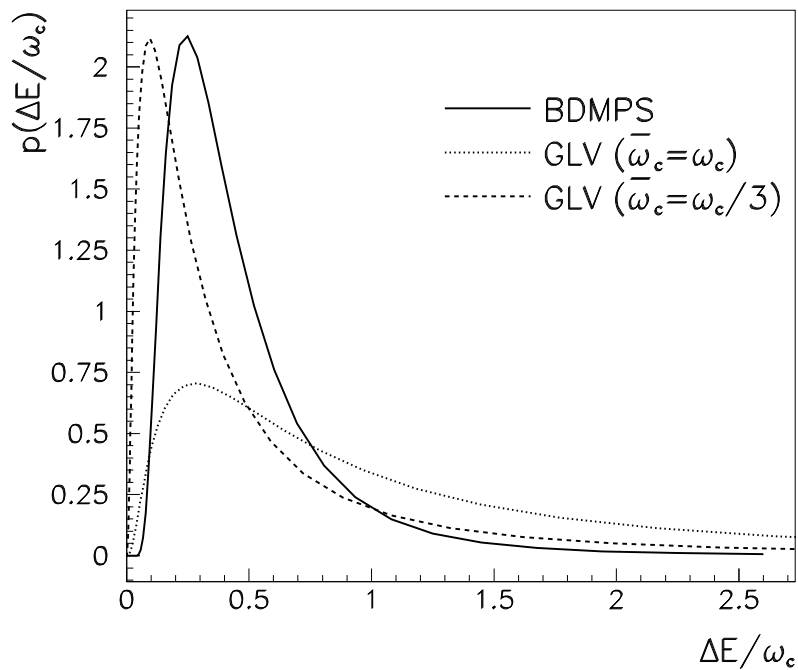}}
\vspace{0.5cm}
\caption{Comparison of the quenching weights for infinite system
size ($R, \bar{R} \to \infty$) calculated for a hard gluon 
in the multiple soft
(``BDMPS'') and single hard (``GLV'') scattering approximation. 
For rescaled characteristic gluon energy,
$\omega_c \simeq (n_0L)\, \bar{\omega}_c$, $n_0L =3$, the
agreement between both probability distributions improves, see
text for further discussion.
}\label{fig10}
\end{figure}

 We start by comparing the $R$- and $\bar{R}$-dependences of the
discrete weight $p_0$ calculated in the multiple soft (Fig.\ref{fig6})
and single hard (Fig.\ref{fig8}) scattering approximation, respectively.
For the choice $R\simeq 3\, \bar{R}$, the curves show better 
agreement. However, the excess above unity for $R< 100$ is much
more pronounced in the multiple soft scattering approximation,
than the excess above unity for $3 \bar{R} < 100$ in the single
hard scattering approximation. This indicates that the specific
destructive interference effects discussed in section \ref{sec3a}
play a more important role in the multiple soft scattering approximation.

In Fig.\ref{fig10}, we compare both approximations in the limit in
which the constraint on the transverse momentum phase space is
removed (i.e. $R, \bar{R} \to \infty$). For the opacity $n_0L = 3$, 
the maximum of the quenching weight takes the same value in both
approximations. However, significant differences can be seen in
the functional shape. The gluon energy distribution is harder
in the single hard scattering approximation (see Fig.\ref{fig5})
and this is reflected in a more pronounced large energy tail of 
the quenching weight.

 We regard the remaining differences between both approximations 
as an indication
of the intrinsic theoretical uncertainties in evaluating the gluon
energy distribution (\ref{2.1}). 

\section{Medium-induced gluon radiation for an expanding medium}
\label{sec4}
Hard partons produced in the initial stage of ultra relativistic
nucleus-nucleus collisions are propagating through a strongly
expanding medium. This results in a time-dependence of the 
transport coefficient $\hat{q}(\xi)$ which can be parametrized 
in terms of a power law,
\begin{equation}
  \hat{q}(\xi) = \hat{q}_0\, \left(\frac{\xi_0}{\xi}\right)^\alpha\, .
  \label{4.1}
\end{equation}
The expansion parameter $\alpha$ determines the dynamical
evolution of the medium: $\alpha = 0$ characterizes a static medium. 
A one-dimensional, boost-invariant
longitudinal expansion is described by $\alpha = 1$. This
value is supported by hydrodynamical simulations of the early stage.
In general, however, an additional transverse expansion can lead to
larger values $\alpha \leq 3$. The maximal value $\hat{q}_0$  of the 
transport coefficient is reached at the time of highest density of
the system which is the formation time $\xi_0$. This formation time
may be set by the inverse of the saturation scale 
$p_{\rm sat}$\cite{Eskola:1999fc}, resulting in $\approx 0.2$ fm/c at
RHIC and $\approx 0.1$ fm/c at LHC. The difference between $\xi_0$
and the production time of the hard parton is negligible for the
calculation of the gluon energy distribution (\ref{2.1}). It will
be ignored in what follows.

In this section, we discuss the range of validity and the form of 
a dynamical scaling law~\cite{Salgado:2002cd}
which relates the gluon energy distribution (\ref{2.1}) in a collision
of arbitrary dynamical expansion to an equivalent static scenario. 

\subsubsection{Multiple soft scattering in an expanding medium}
\label{sec4a}
In Appendix \ref{appc}, we give details of the calculation of the
gluon energy distribution (\ref{2.1}) for values of the expansion
parameter $\alpha < 3$ in the multiple soft scattering 
approximation (\ref{2.3}).
As reported previously~\cite{Salgado:2002cd}, we observe a scaling law
which relates the time-dependent transport coefficient (\ref{4.1}) to
an equivalent static transport coefficient $\bar{\hat{q}}$,
\vspace{-0.5cm}
\begin{figure}[h]\epsfxsize=14.7cm
\centerline{\epsfbox{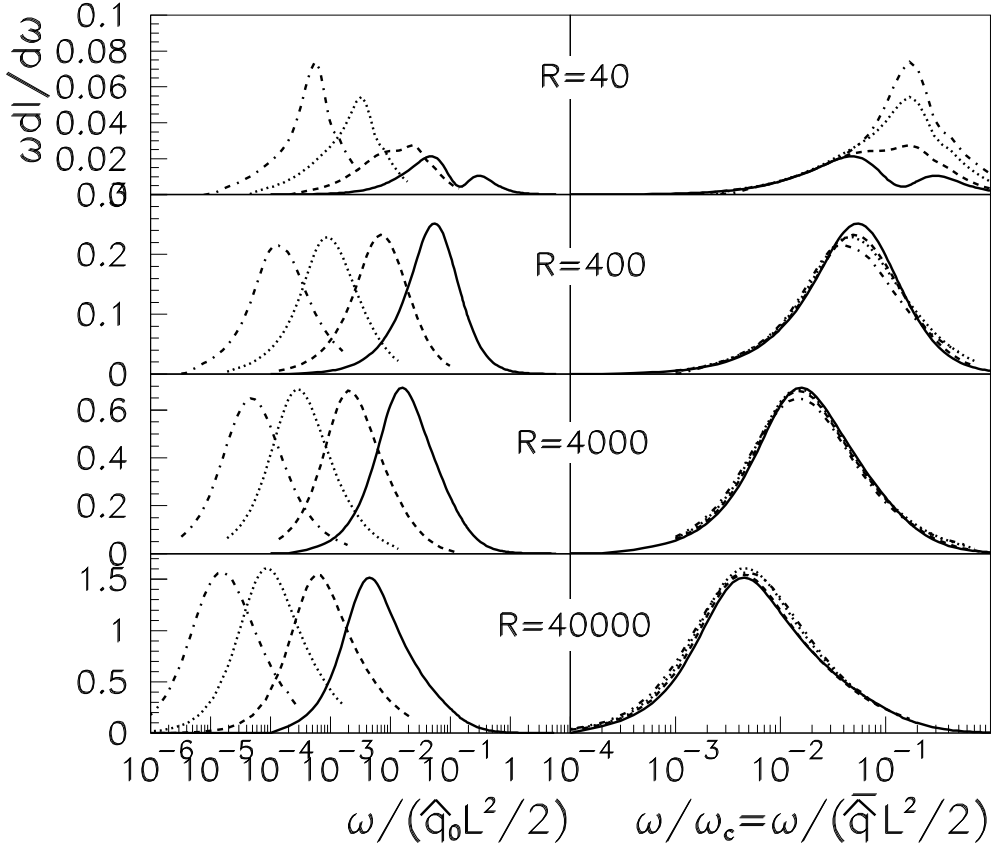}}
\vspace{0.5cm}
\caption{The gluon energy distribution calculated in the multiple
soft scattering approximation with expansion parameter $\alpha =0$
(solid line), $\alpha = 0.5$ (dashed line), $\alpha = 1.0$
(dotted line) and $\alpha = 1.5$ (dash-dotted line). Curves in the
right hand column are scaled according to (\protect\ref{4.2}).
}\label{fig11}
\end{figure}
%
\begin{eqnarray}
  \bar{\hat{q}} = \frac{2}{L^2} \int_{\xi_0}^{L+\xi_0} d\xi\, 
  \left(\xi - \xi_0\right)\, \hat{q}(\xi)\, .
  \label{4.2}
\end{eqnarray}
As seen in Fig.~\ref{fig11}, the gluon energy distributions for 
different values of the expansion parameter $\alpha$ differ by
orders of magnitude if plotted in units of the same characteristic
gluon energy $\omega_c = \frac{1}{2} \hat{q}_0\, L^2$ and kinematic
constraint $R = \omega_c\, L$. However, if plotted in units of the
rescaled gluon energy $\frac{1}{2} \bar{\hat{q}}\, L^2$ and the
rescaled kinematic constraint $\frac{1}{2} \bar{\hat{q}}\, L^3$,
they agree approximately over a large parameter range.

For practical purposes, the accuracy of the scaling law (\ref{4.2}) 
is satisfactory for $R > 100$. Concerning the deviations from the 
scaling law for $R < 100$ (see Fig.\ref{fig11}), we make the
following comments: 
In practice, these deviations are negligible since $p_0 \sim 1$
for $R < 100$ and thus no significant medium modification occurs.
Technically, the static case ($\alpha = 0$) is calculated for
a box profile in the longitudinal density of scattering centers. 
On the other hand, in the expanding scenarios, 
the density profile degrades more smoothly with increasing path length,
and the discontinuity at path length $L$ is less important. The strength
of destructive interferences between medium-induced and vacuum gluon
radiation depends on this discontinuity. This may explain why for
$R = 40$ the
rescaled gluon energy distribution in Fig.~\ref{fig11} is more suppressed
in the static case than in the expanding case.

\subsubsection{The opacity expansion for an expanding medium}
\label{sec4b}
In Appendix \ref{appb}, we give analytical expression for the 
single hard scattering limit of the gluon energy distribution (\ref{2.1})
in a medium with expansion parameter $\alpha$. The analytical form 
of (\ref{2.1}) changes with the expansion parameter $\alpha$. 
We derive an explicit expression for the Bjorken scaling case 
$\alpha = 1$, 
\begin{eqnarray}
  \omega \frac{dI_{\alpha=1}^{N=1}}{d\omega} 
   &=& 2 \frac{\alpha_s\, C_R}{\pi}\,
   (n_0\xi_0)\,   
  \int_0^\infty \frac{dr}{r}\,
  {\rm Re}\left[- {\rm Ei}[-ir] + \ln [-ir] + \gamma_E \right]
 \nonumber \\
 && \qquad \times
  \left( \frac{\gamma}{r + \gamma} - 
         \frac{\gamma}{\sqrt{(\kappa^2 + r + \gamma)^2
                       - 4 \kappa^2 r}}\right)\, .
  \label{4.3}
\end{eqnarray}
Here, $\gamma_E = 0.577 \dots$ denotes Euler's constant and
the exponential integral function is
${\rm Ei}[z] = - \int_{-z}^\infty dt\, e^{-t}/t$.

\begin{figure}[h]\epsfxsize=14.7cm
\centerline{\epsfbox{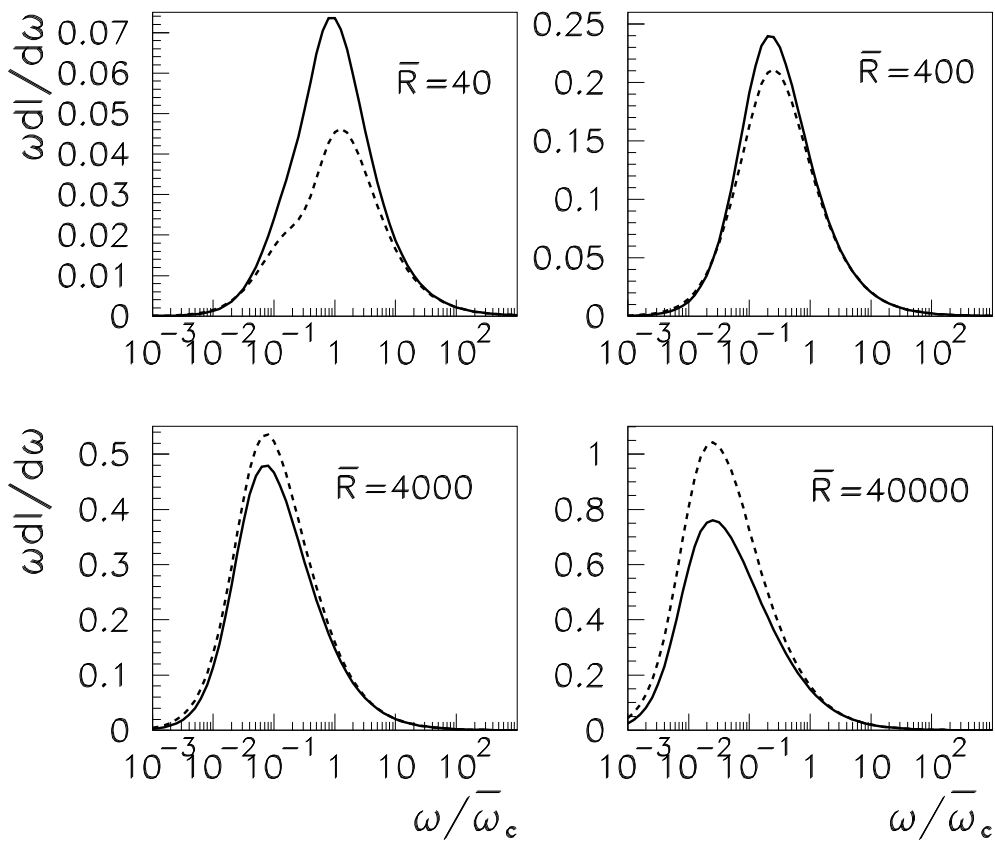}}
\vspace{0.5cm}
\caption{The gluon energy distribution calculated in the single
hard scattering approximation for a static medium (dashed line)
and for a medium with Bjorken expansion (solid line). Curves for
the dynamically expanding case are scaled according to
eq. (\protect\ref{4.4}).
}\label{fig12}
\end{figure}

To relate the gluon energy distributions for a static medium
(\ref{2.19}) and a Bjorken scaling expansion (\ref{4.3}), 
we determine the dynamically averaged density of scattering
centers following eq. (\ref{4.2})
\begin{equation}
  \bar{n} = \frac{2}{L^2} \int_{\xi_0}^{L+\xi_0} d\xi\, 
  \left(\xi - \xi_0\right)\, n(\xi)
  = \frac{2\, n_0\, \xi_0}{L}\, .
  \label{4.4}
\end{equation} 
This equation suggests that the gluon energy distributions in the static
and Bjorken expansion case show agreement if the prefactor $(n_0\xi_0)$ 
in eq. (\ref{4.3}) is replaced by $\frac{1}{2} \bar{n}\, L$  
where $\bar{n}$ determines the density of scattering centers of the 
equivalent static scenario. In Fig.\ref{fig12}, we test this
suggestion numerically for different values of the kinematic
constraint $\bar{R}$.

Remarkably, for sufficiently large kinematic constraint 
$\bar{R} > 100$, the gluon energy distribution shows dynamical
scaling according to (\ref{4.4}) for gluon energies 
$\omega > \bar{\omega}_c$. This is the dominant kinematic 
region in the opacity expansion (see the discussion in
section ~\ref{sec2b}, eq. (\ref{2.21}) ff). Thus,
despite the deviations from the scaling law for 
$\omega < \bar{\omega}_c$, the logarithmically enhanced
contribution to the average energy loss
\vspace{-0.5cm}
%
\begin{equation}
  \langle \Delta E \rangle^{N=1} 
  = \int d\omega\, \omega\, \frac{dI^{N=1}}{d\omega}\quad
  \hbox{shows scaling\protect\cite{Gyulassy:2000gk} with (\protect\ref{4.4}).}
  \label{4.5}
\end{equation}
This is also known to hold in the soft multiple scattering 
approximation~\cite{Baier:1998yf} and it is consistent with
results obtained on the basis of twist-4 matrix 
elements \cite{Wang:2002ri}.

In the region $\omega < \bar{\omega}_c$, significant deviations
from the scaling law (\ref{4.4}) are seen in Fig.~\ref{fig12}.
However, the logarithmic plot overemphasizes the importance of
these deviations. First, they occur in the sub dominant region
which is less important for calculating the quenching weights.
Second, these deviations do not exceed 30 percent in the physically
relevant parameter range $100 < \bar{R} < 40000$ in which significant
medium modifications can be expected.

For practical purposes, the scaling law (\ref{4.4}) is thus
satisfactory. Quenching weights for a dynamically expanding
scenario can be obtained by calculating the quenching weights
of the dynamically equivalent static scenario according to
(\ref{4.4}).

\section{Angular dependence of radiation probability}
\label{sec5}
The maximal angle under which a gluon can be radiated is given by 
the upper bound on the transverse momentum integral in (\ref{2.1}),
\begin{equation}
  \Theta \simeq \frac{k_\perp^{\rm max}}{\omega} = \chi\, .
  \label{5.1}
\end{equation}
Thus, for fixed values of the characteristic gluon energy $\omega_c$
and of the kinematic constraint $R=\omega_c\, L$, a decreasing value of
$R_{\chi} = \chi^2 \omega_c\, L$ determines the medium-induced energy
radiated into a cone of opening angle $\Theta$. In this section,
we denote explicitly the dependence of the quenching weight
$P(\Delta E, \omega_c, R_{\chi})$ on $\omega_c$ and $\chi^2\, R$. 
This quenching weight determines the probability that an additional 
energy fraction $\Delta E$ is radiated {\it inside} the opening 
angle $\chi = \Theta$. From the Figs. \ref{fig1} and \ref{fig3}, 
we know that the more collinear component of the medium-induced 
gluon radiation is harder.

For fixed values of $\omega_c$ and $R=\omega_c\, L$, the gluon energy
distribution radiated {\it outside} the opening angle $\Theta$ is
given by
\begin{equation}
  \omega \frac{dI^{>\Theta}}{d\omega}(\omega_c,R) = 
  \omega \frac{dI}{d\omega}(\omega_c,R) 
  - \omega \frac{dI}{d\omega}(\omega_c,R_{\chi})\, .
  \label{5.2}
\end{equation}
The probability that an additional energy fraction $\Delta E$ is
radiated {\it outside} the opening angle $\chi \simeq \Theta$ is obtained
by inserting (\ref{5.2}) into the Mellin transform (\ref{3.2}),
(\ref{3.3}). For the current work, we did not calculate this
probability; there is no simple way to obtain it without
Mellin transform directly from the quenching weights tabulated
in section~\ref{sec3}.

\begin{figure}[h]\epsfxsize=16.7cm
\centerline{\epsfbox{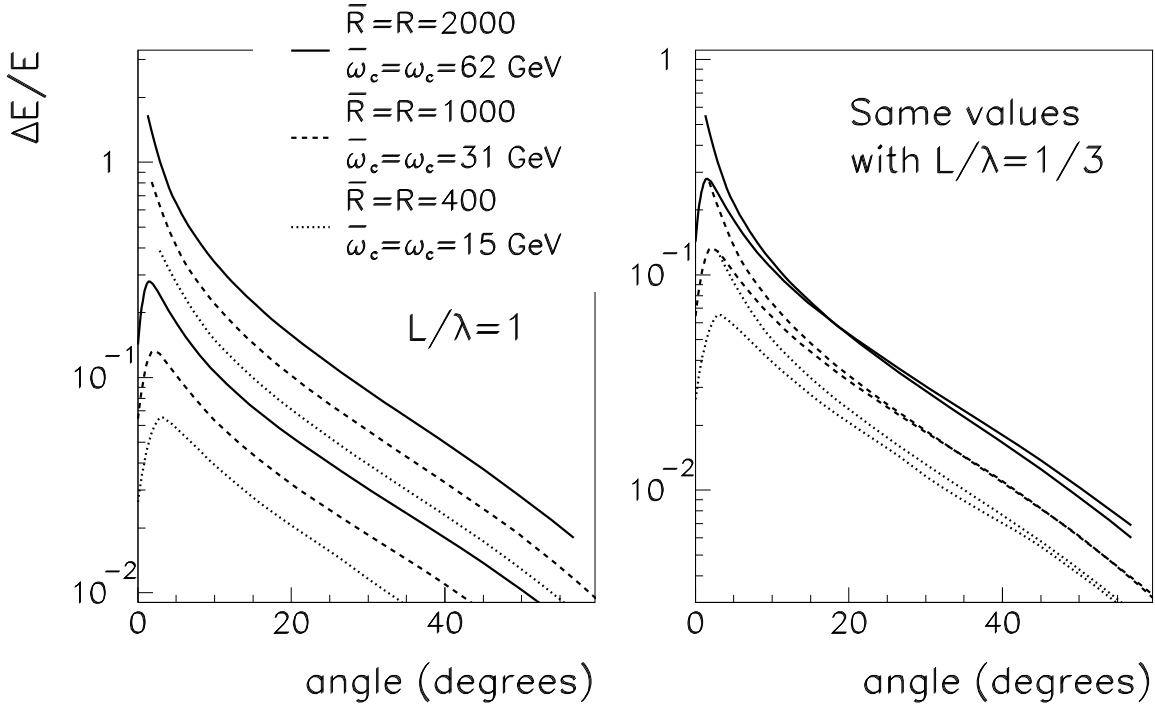}}
\vspace{0.5cm}
\caption{The average energy loss (\protect\ref{5.3}) radiated
outside an angle $\Theta$ as calculated in the multiple soft
(lower three lines) and single hard (upper three lines) 
scattering approximation for a jet of energy $E= 100$ GeV.
}\label{fig13}
\end{figure}

The calculation of the average energy loss {\it outside}
an angle $\Theta$ is simpler.
It can be calculated from the quenching weights 
tabulated in section~\ref{sec3} 
\begin{eqnarray}
  \langle \Delta E \rangle(\Theta) 
  &=& \int d\omega\, \omega\, 
   \frac{dI^{>\Theta}}{d\omega}(\omega_c,R=\omega_cL)
   \nonumber \\
  &=& \int d\bar{E}\, \bar{E}\, 
  \left[ P(\bar{E},\omega_c,R=\omega_cL) 
         - P(\bar{E},\omega_c,R_\chi=\chi^2\omega_cL) 
         \right]\, .
  \label{5.3}
\end{eqnarray}
In Fig.\ref{fig13}, we compare the angular dependence of the
average parton energy loss (\ref{5.3}) in the multiple soft
and single hard scattering approximation. 

In the single hard scattering approximation, the integral
(\ref{5.3}) diverges logarithmically in the ultraviolet for
$\chi = \Theta \to 0$. For the calculation of 
$\langle \Delta E\rangle^{N=1}(\Theta = 0)$ in (\ref{2.22}),
we have cut off this divergence by limiting the energy 
radiated away to $\omega < E$. For the plot in Fig.\ref{fig13},
we restrict instead the calculation to sufficiently large 
angles $\Theta$ for which the second term in (\ref{5.3})
provides an ultraviolet cut-off. Thus, for small angles $\Theta < 10^\circ$
where $\Delta E \sim E$, one
overestimates $\langle \Delta E\rangle^{N=1}(\Theta)$.
\vspace{-0.5cm}

In the single
hard scattering approximation, the region $\omega > \bar{\omega}_c$
is dominant. This hard part of the spectrum is emitted under angles
$\Theta < \mu/\bar{\omega}_c$ and thus appears as a logarithmic
enhancement in the collinear region. In the multiple soft scattering
approximation, however, the dominant radiative contribution lies in
the soft region $\omega < \omega_c$ which is emitted under relatively
large opening angles $\Theta > k_\perp/\omega_c \sim
\hat{q}^{1/4}/\omega_c^{3/4}$. For smaller opening angles,
the average energy loss $\langle \Delta E\rangle(\Theta)$ does not
increase further. Indeed, multiple soft scattering results in
a shift in transverse phase space which is known to deplete
$\langle \Delta E\rangle(\Theta)$ at very small 
angles\cite{Wiedemann:2000tf,Baier:2001qw}. Thus, in 
the multiple soft scattering approximation, there is no ultraviolet
divergence at small angle $\Theta$.

To compare the single hard and multiple soft scattering approximations
for $\langle \Delta E\rangle(\Theta)$, we proceed in analogy to
the discussion in section \ref{sec3c}: varying the opacity, we
find the best agreement between both approximations for
$n_0\, L = 3$. A qualitative difference which cannot be 
adjusted by the choice of the additional model parameter $n_0\, L$
persists for small angles only. Its origin is explained above.
Thus, Fig.\ref{fig13} indicates that for comparable sets of model
parameters $\omega_c$, $R$ and $\bar{\omega}_c$, $\bar{R}$, 
$n_0\, L$ respectively, the multiple soft and single hard 
scattering approximations lead to a comparable angular 
dependence of $\langle \Delta E\rangle(\Theta)$ for $\Theta > 10^{\circ}$. 
\vspace{-0.5cm}
\begin{figure}[h]\epsfxsize=10.7cm
\centerline{\epsfbox{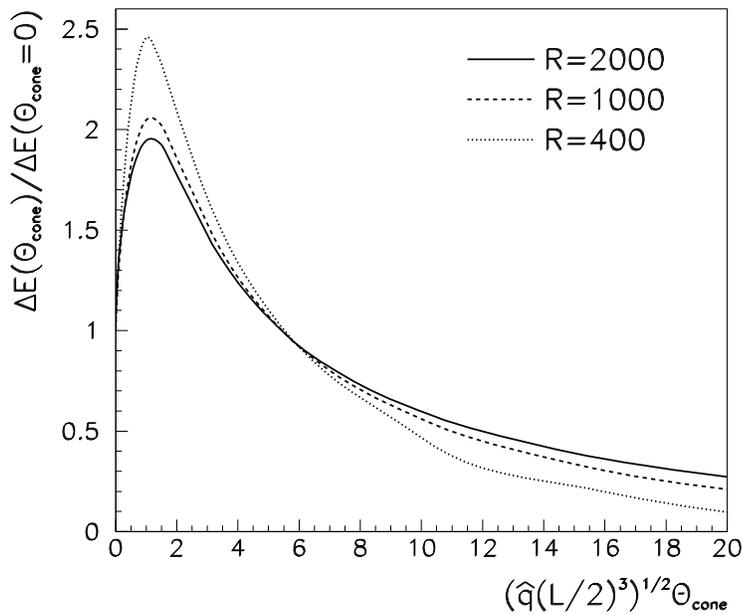}}
\vspace{0.5cm}
\caption{Angular dependence of the average energy loss 
(\protect\ref{5.3}) in the multiple soft scattering approximation
as a function of the rescaled parameter 
$\hat{q} (L/2)^3\, \Theta$.
}\label{fig14}
\end{figure}
%
The calculation of the angular dependence in Fig.\ref{fig13} was
done for quenching weights calculated for a static medium. In
general, the dynamical scaling laws (\ref{4.2}) and (\ref{4.4})
relate these to the quenching weights of dynamically expanding 
scenarios. However, this is not the case for the small values
$R$, $\bar{R} < 100$ which enter the calculation of the small
angular dependence ($\Theta < 10^\circ$) in eq. (\ref{5.3}). 
For these small values of $R$, $\bar{R}$, dynamical scaling breaks
down (see Figs.\ref{fig11} and \ref{fig12}). Since typical jet
cone openings correspond to larger angles $\Theta > 10^\circ$,
we did not make an effort to quantify the remaining dependence of
$\langle \Delta E\rangle(\Theta)$ on the collective expansion of
the collision region.

In the multiple soft scattering approximation, 
$\langle \Delta E\rangle(\Theta=0)$ is finite. Baier et al.
observed\cite{Baier:1999ds} that the ration 
$\langle \Delta E\rangle(\Theta)/ \langle \Delta E\rangle(\Theta=0)$
is an universal quantity which depends solely on $\hat{q}\, L^3\,\Theta$.
Fig. \ref{fig14} shows that this statement remains 
approximately true in the presence of a finite kinematic
constraint $R$.

\section{Applications of quenching weights}
\label{sec6}

In this section, we use quenching weights to calculate in two alternative
ways the medium-induced suppression of hadronic high transverse momentum 
spectra. In section~\ref{sec6a}, we determine the quenching factor
$Q(p_\perp)$ and in section~\ref{sec6b} we calculate medium-modified
parton fragmentation functions. In section~\ref{sec6c} we finally
discuss the relation of both calculations to the nuclear modification 
factor measured at RHIC.

\subsection{Quenching factors for hadronic spectra}
\label{sec6a}
The medium-dependence of inclusive transverse momentum
spectra can be characterized in terms of the quenching 
factor\cite{Baier:2001yt}
\begin{eqnarray}
 Q(p_\perp)&=&
{{d\sigma^{\rm med}(p_\perp)/ dp^2_\perp}\over
{d\sigma^{\rm vac}(p_\perp)/ dp^2_\perp}}=
\int d{\Delta E}\, P(\Delta E)\left(
{d\sigma^{\rm vac}(p_\perp+\Delta E)/ dp^2_\perp}\over
{d\sigma^{\rm vac}(p_\perp)/ dp^2_\perp}\right)\, .
 \label{6.1}
\end{eqnarray}
Here, the spectrum 
${d\sigma^{\rm vac}(p_\perp)/ dp^2_\perp}$ is unaffected by 
medium effects; it is determined e.g. in proton-proton collisions.
Equation (\ref{6.1}) relates it to the medium-modified transverse 
momentum spectrum ${d\sigma^{\rm med}(p_\perp)/ dp^2_\perp}$ 
measured e.g. in nucleus-nucleus collisions.
We work in the longitudinally comoving frame in which the
total energy of the produced parton is directed orthogonal
to the beam. Due to the presence of the medium, a parton 
produced initially with transverse momentum $p_\perp+\Delta E$ 
looses an additional energy $\Delta E$ with probability
$ P(\Delta E)$. This defines the quenching factor (\ref{6.1}).

If one assumes a power law fall-off of the $p_\perp$-spectrum,
then the quenching factor (\ref{6.1}) can be calculated
explicitly,
\begin{eqnarray}
 Q(p_\perp) &\simeq&  \int d{\Delta E}\, P(\Delta E)\,  
    \left({p_\perp\over p_\perp+\Delta E}\right)^n\, .
 \label{6.2}
\end{eqnarray}
In general, the effective power $n$ depends on $p_\perp$ and
$\sqrt{s}$. It is $n \simeq 7$ in the kinematic range relevant
for RHIC. 
%
\begin{figure}[h]\epsfxsize=14.7cm
\centerline{\epsfbox{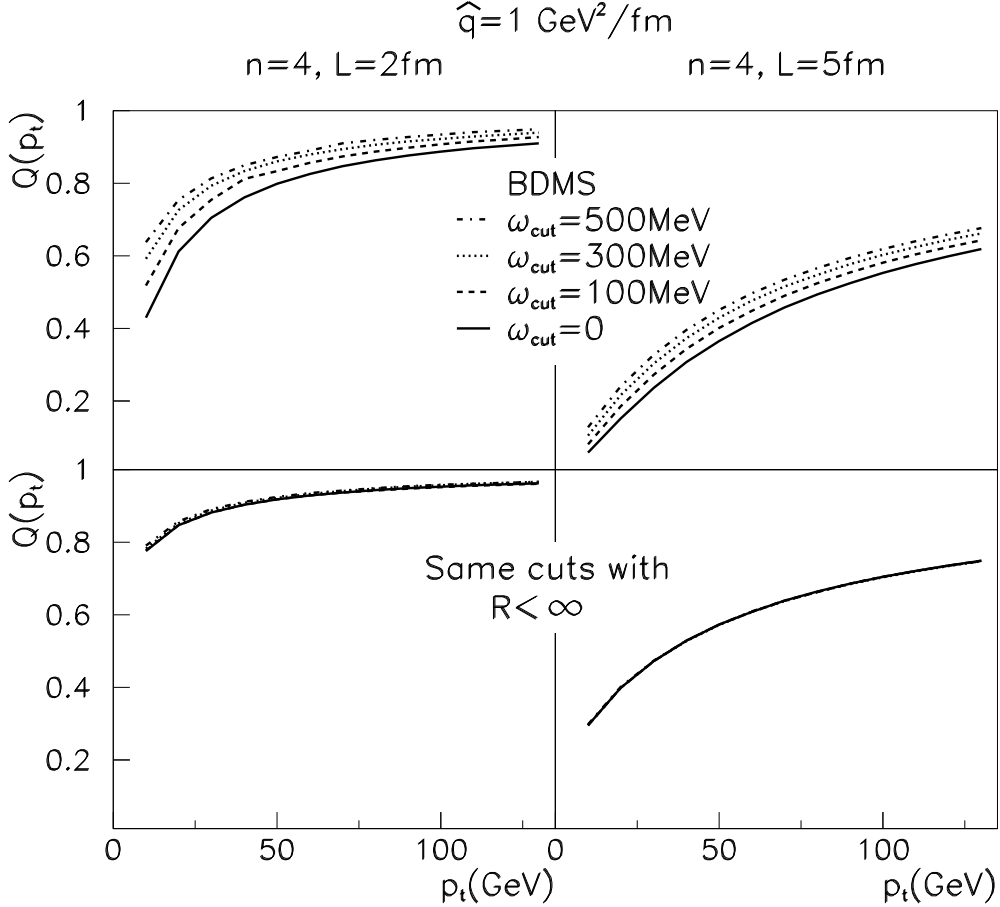}}
\caption{The quenching factor (\protect\ref{6.2}) calculated in the
soft multiple scattering approximation for $\alpha_s = 1/2$. 
Upper row: calculation in the 
$R\to\infty$-limit but with a varying sharp cut-off on the 
infrared part of the gluon energy distribution $\omega \frac{dI}{d\omega}$
determining the quenching weight. Lower row: the 
same calculation is insensitive to infrared contributions if 
the finite kinematic constraint $R=\omega_cL <\infty$ is included. 
}\label{fig15}
\end{figure}
%
To compare directly to published results\cite{Baier:2001yt}, 
we calculate the quenching factor (\ref{6.2}) in Fig. \ref{fig15}
for parameter values used previously. The transport coefficient 
is taken to match expectations for a hot medium, 
$\hat{q} = \frac{(1 {\rm GeV})^2}{\rm fm}$. Given the in-medium
path length $L$, this defines $\omega_c$ and $R$ in the multiple
soft scattering approximation. In the single hard scattering
approximation, parameters are chosen for opacity
$n_0\, L = 1$ by identifying $\bar{R}=R$ and 
$\bar{\omega}_c = \omega_c$. The effective power $n$ in 
(\ref{6.2}) is set to its asymptotic value $n=4$. To be 
quantitatively comparable with Ref. \cite{Baier:2001yt}, we
use $\alpha_s = 1/2$ in Figs. \ref{fig15} and \ref{fig16}
while all other numerical results are given for $\alpha_s = 1/3$.

Fig. \ref{fig15} shows the numerical results obtained in the
multiple soft scattering approximation. The perturbative calculation
of the gluon energy distribution $\omega \frac{dI}{d\omega}$ cannot
be trusted for soft gluon energies $\omega \sim O(\Lambda_{\rm QCD})$
where perturbation theory breaks down. 
To quantify the sensitivity of their calculation to this infrared
region, Baier et al.\cite{Baier:2001yt} introduced a sharp 
cut-off on the $R\to \infty$ gluon energy distribution which was
varied between $\omega_{\rm cut} = 0$ and $\omega_{\rm cut} = 500$ MeV.
The resulting uncertainty is seen as a $\approx$ 20\% variation
of the quenching factor $Q(p_\perp)$ in the upper row of
Fig.\ref{fig15}. The finite kinematic 
constraint $R=\omega_cL <\infty$ depletes the infrared region of 
the medium-induced gluon radiation spectrum (see Fig. \ref{fig1}). 
This constraint is a generic consequence of a finite in-medium path 
length; it
removes almost completely the sensitivity of the calculation
to the uncontrolled infrared region (see Fig.\ref{fig15}, lower row).
Remarkably, it also tends to flatten the $p_\perp$-dependence of the
quenching factor. We shall return to the consequences of this
observation when we zoom into the region $p_\perp < 10$ GeV
in section~\ref{sec6c}.
%
\begin{figure}[h]\epsfxsize=14.7cm
\centerline{\epsfbox{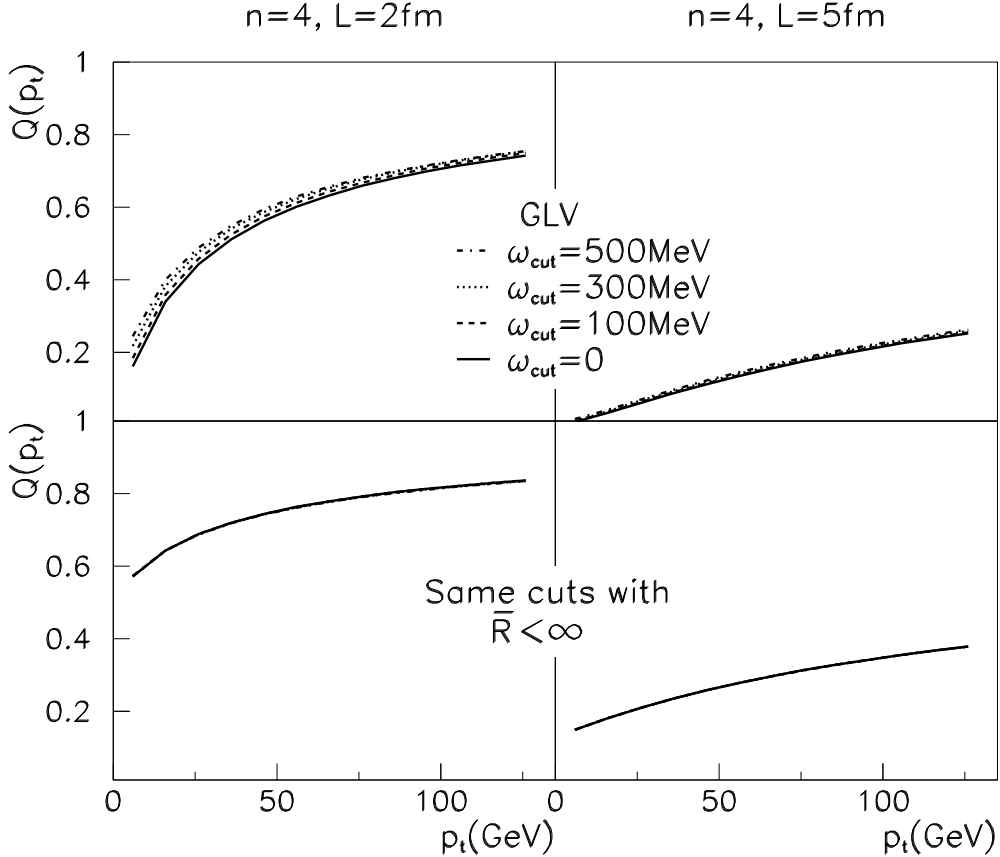}}
\caption{The same quenching factor (\protect\ref{6.2}) as in
Fig.\protect\ref{fig15}, here calculated in the single hard
scattering approximation. 
}\label{fig16}
\end{figure}
%
We have calculated the quenching factor (\ref{6.1}) in the single 
hard scattering approximation, see Fig. \ref{fig16}. 
Since the dominant contribution comes in this case from the hard
part of the spectrum, $\omega > \bar{\omega}_c$, the
sensitivity to the infrared cut $\omega_{\rm cut}$ is
much reduced in comparison to Fig. \ref{fig15}. However,
realistic finite kinematic constraints $\bar{R}$ remove a
much larger part of the soft spectrum. As a consequence, the 
absolute value of $Q(p_\perp)$ increases significantly if
finite kinematic constraints are imposed, and the 
$p_\perp$-dependence tends to flatten.

\begin{figure}[h]\epsfxsize=10.7cm
\centerline{\epsfbox{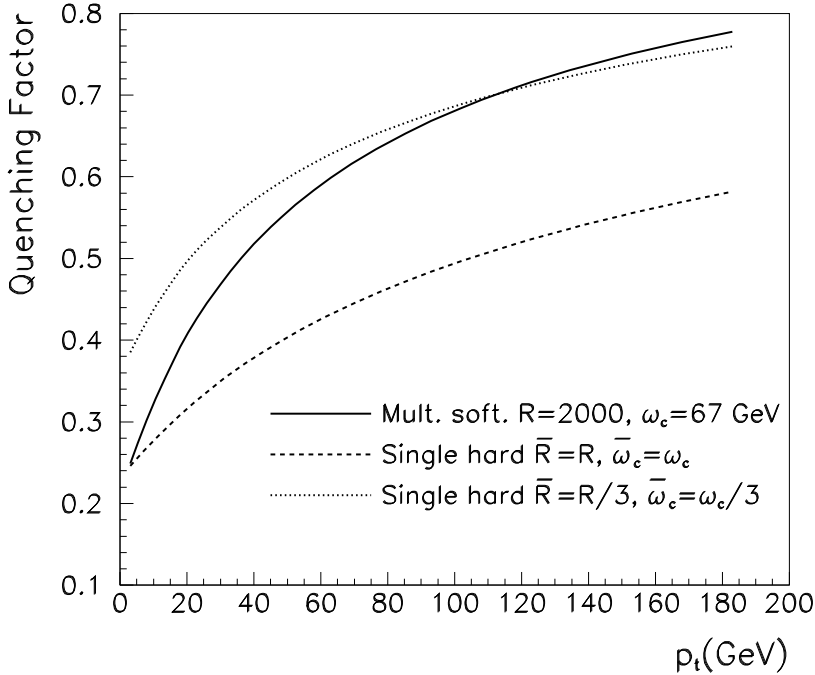}}
\vspace{0.5cm}
\caption{Comparison of the quenching factor (\protect\ref{6.2})
calculated for $\alpha_s = 1/3$ and effective power $n=7$ in
the multiple soft, single hard and rescaled single hard
scattering approximation.
}\label{fig17}
\end{figure}

In section \ref{sec3c}, we observed that for quenching
weights the best agreement between single hard and
multiple soft scattering approximation is for 
$\omega_c =  3\bar{\omega}_c$, $R = 3 \bar{R}$, $n_0L = 3$.
Remaining differences come from the fact that the single
hard scattering approximation shows a dominant contribution
for $\omega > \bar{\omega}_c$ while the multiple soft scattering
approximation shows a dominant contribution for $\omega < \omega_c$.
However, the quenching factor $Q(p_\perp)$ for small $p_\perp$ is 
sensitive only to the soft region $\omega < \omega_c$,
($\omega < \bar{\omega}_c$) in {\it both} approximations. This
is so because medium-induced gluons cannot carry away more 
than the total energy $E_q$ of the parent parton, and hence
$\omega < E_q < \omega_c$ ($\bar{\omega}_c$) at small $p_\perp$. 
Thus, the simple relation $\omega_c =  3\bar{\omega}_c$, $R = 3 \bar{R}$
does not hold for $p_\perp < \omega_c$. This is seen in
Fig. \ref{fig17}. For small $p_\perp$, the multiple soft scattering 
approximation results in a much stronger suppression than 
the single hard one calculated for rescaled parameters 
$\omega_c =  3\bar{\omega}_c$, $R = 3 \bar{R}$.
%

\subsection{Medium-modified fragmentation functions}
\label{sec6b}

Medium-induced gluon radiation off hard partons modifies the 
fragmentation and hadronization of final state partons, thus
affecting hadronic $p_\perp$-spectra. In section~\ref{sec6a},
we calculated this effect in terms of the quenching factor $Q(p_\perp)$.
Alternatively, this quenching factor can be determined from
medium-modified fragmentation functions, which we discuss now.

In the QCD-improved parton model, hadronic cross sections
for high-$p_\perp$ hadroproduction are calculated by convoluting 
the perturbatively calculable hard partonic cross section
$d\sigma^q$
and the (final state) fragmentation function $D_{h/q}(x,Q^2)$,
\begin{equation}
  d\sigma^h(z,Q^2) = \left( \frac{d\sigma^q}{dy}\right) dy\,
  D_{h/q}(x,Q^2)\, dx\, . 
  \label{6.3}
\end{equation}
For notational simplicity, we do not denote the additional convolution 
of $d\sigma^q$ with the (initial state) parton distributions.
The leading hadron $h$ carries an energy fraction 
$z = E_h/Q$ of the total virtuality of the partonic collision,
which is a fraction $x=E_h/E_q$ of the energy of its parent parton. 
The parent parton carries the energy fraction $y=E_q/Q$.

If the parent parton loses with probability $P(\epsilon)$
an additional energy fraction $\epsilon = \frac{\Delta E}{E_q}$ 
prior to hadronization, then the leading hadron is a fragment of
a parton with lower energy $(1-\epsilon) E_q$; thus, it carries
a larger fraction $\frac{x}{1-\epsilon}$ of the initial parton energy. 
The inclusion of this effect amounts to replacing
the fragmentation function $D_{h/q}(x,Q^2)$ in (\ref{6.3}) 
by the medium-modified
fragmentation function~\cite{Wang:1996yh,Gyulassy:2001nm}
\begin{eqnarray}
  D_{h/q}^{(\rm med)}(x,Q^2) 
  = \int_0^1 d\epsilon\, P(\epsilon)\,
  \frac{1}{1-\epsilon}\, 
  D_{h/q}(\frac{x}{1-\epsilon},Q^2)\, .
  \label{6.4}
\end{eqnarray}
To calculate eq. (\ref{6.4}), we use the recent LO fragmentation functions 
of Kniehl, Kramer and P\"otter \cite{Kniehl:2000fe} (``KKP'').
These improve over previously available 
parametrizations\cite{Binnewies:1994ju}. However, the KKP parametrization
still shows significant 
uncertainties in the large-$x$ region relevant for hadronic 
$p_\perp$-spectra \cite{Zhang:2002py}.
For alternative approaches towards medium-modified fragmentation
functions, see Refs.\cite{Guo:2000nz,Osborne:2002dx}.
%
\begin{figure}[h]\epsfxsize=12.7cm
\centerline{\epsfbox{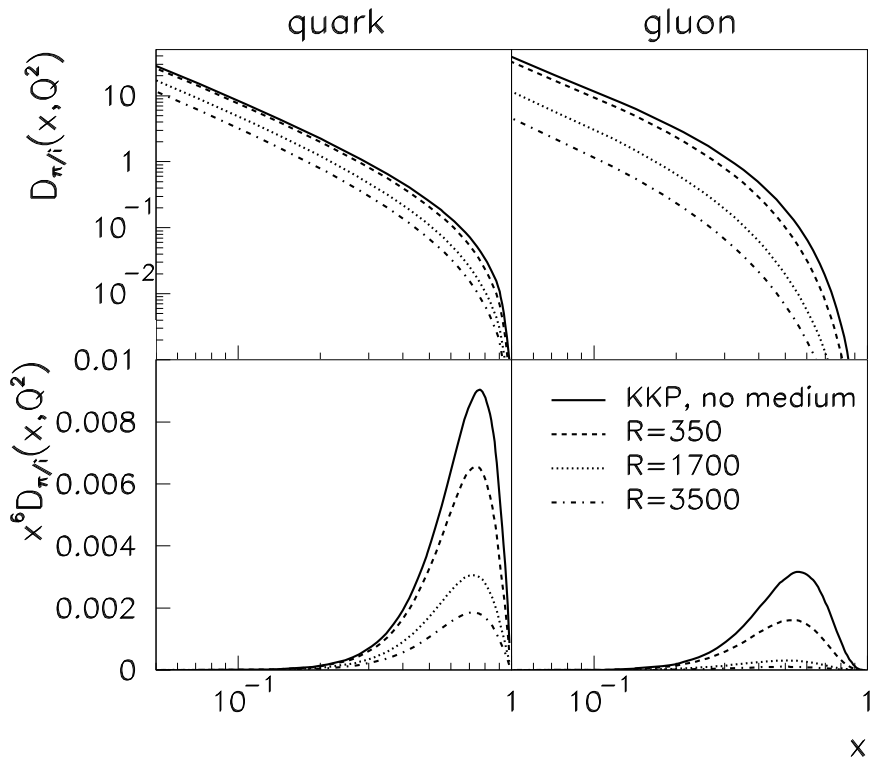}}
\caption{The medium-modified fragmentation function (\protect\ref{6.4})
for $Q^2 = (10\, {\rm GeV})^2$
calculated in the multiple soft scattering approximation for a medium
of length $L = 6$ fm.
}\label{fig18}
\end{figure}
%
We have calculated the medium-modified $q\to\pi$ and $g\to\pi$
fragmentation functions (\ref{6.4}), using the quenching weights
in the multiple soft (Fig.\ref{fig18}) and single hard (Fig.\ref{fig19})
scattering approximation. The energy of the parent parton is set to the 
virtuality of the hard process \cite{Salgado:2002cd}, $E_q \sim Q$. 
The medium-induced fragmentation functions decrease with
increasing density of the medium since the probability of a 
parton of initial energy $E_q$ to
fragment into a hadron of large energy $xE_q$ decreases with
increasing parton energy loss.
They should be trusted for sufficiently large momentum fractions
($x > 0.1$ say) only. The reason is that the hadronized remnants
of the medium-induced soft radiation are not included in the definition
of (\ref{6.4}). These remnants are soft - they can be expected
to give an additional contribution in the region $x < 0.1$. The
neglect of these remnants in (\ref{6.4}) implies that the normalization 
of $D_{h/q}^{(\rm med)}(x,Q^2)$ is a factor 
$\int d\epsilon\, \epsilon\, P(\epsilon)$ too small,
\begin{equation}
  \int_0^1 dx\, x\, D_{h/q}^{(\rm med)}(x)
  \simeq \int_0^1 dx\, x\, D_{h/q}(x)
    \int d\epsilon\, (1-\epsilon)\, P(\epsilon)\, .
    \label{6.6}
\end{equation}
For the suppression of high-$p_\perp$ hadronic spectra, 
this normalization error is unimportant since the main
contribution comes from the region of larger $x$.
%
\begin{figure}[h]\epsfxsize=12.7cm
\centerline{\epsfbox{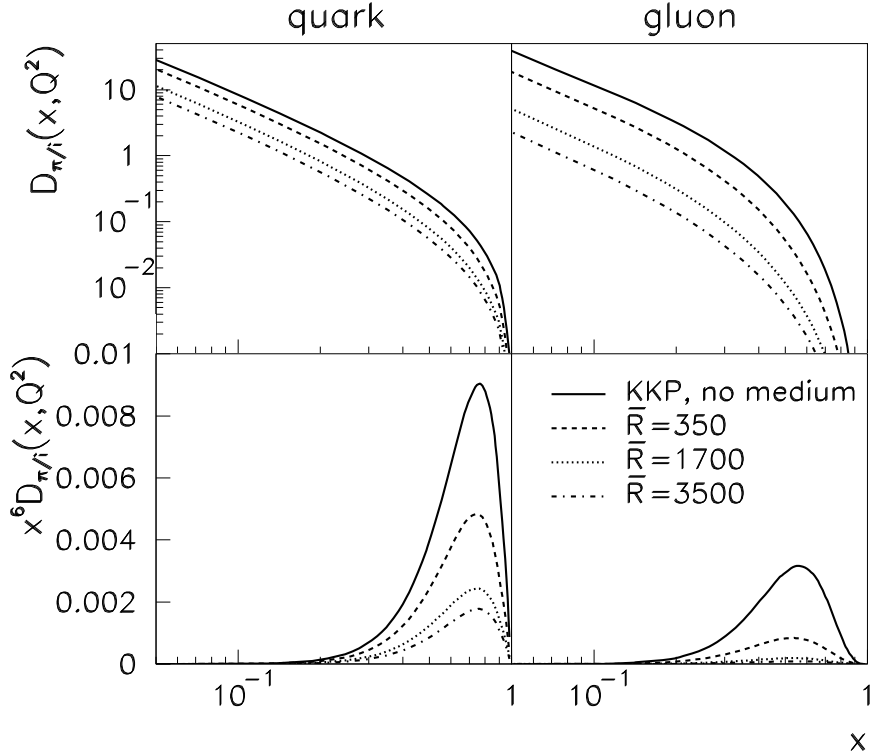}}
\caption{The medium-modified fragmentation function (\protect\ref{6.4})
for $Q^2 = (10\, {\rm GeV})^2$
calculated in the single hard scattering approximation for a medium
of length $L = 6$ fm.
}\label{fig19}
\end{figure}
%
To estimate the corresponding suppression
of high-$p_\perp$ hadronic spectra, we exploit that the fragmentation
function in (\ref{6.3}) is weighed by the partonic cross section
$d\sigma^q/dp_\perp^2$ (we work in the frame in which the total energy 
of the produced parton goes transverse to the beam). In the kinematic
range relevant for RHIC ($\sqrt{s} = 200$ GeV and $p_\perp \sim 10$ GeV),
one finds \cite{Eskola:2002kv} 
${d\sigma^q}/{dp_{\perp}^2} \sim {1}/{p_{\perp}^{n(\sqrt{s}, p_{\perp})}}$
with $n(\sqrt{s}, p_{\perp}) \sim 6$. Thus, eq. (\ref{6.3}) 
effectively tests $x^{n(\sqrt{s}, p_{\perp})} D_{h/q}^{(\rm med)}(x,Q^2)$.
The suppression factor 
\begin{equation}
  R_{ff}(p_\perp)
  = \frac{ x_{\rm max}^6 D_{h/q}^{(\rm med)}(x_{\rm max},p_\perp^2)}{
                   x_{\rm max}^6 D_{h/q}(x_{\rm max},p_\perp^2)}
  \Bigg\vert_{p_\perp = x_{\rm max}\, E_q}
  \label{6.5}
\end{equation}
provides a simple estimate of the reduction of hadronic $p_\perp$-spectra.
In eq. (\ref{6.5}), $x_{\rm max}$ denotes the maximum of 
$x^{n(\sqrt{s}, p_{\perp})} D_{h/q}^{(\rm med)}(x,Q^2)$ and corresponds
to the most likely energy fraction $p_\perp = x_{\rm max}\, E_q$ of the
leading hadron. The suppression factor can be read off easily from the 
lower rows of Figs.\ref{fig18} and \ref{fig19}. We now compare
this suppression factor to the quenching factor $Q(p_\perp)$ in
(\ref{6.2}).

\subsection{The nuclear modification factor}
\label{sec6c}

{\it Experimental situation:}
Published data for Au+Au collisions at $\sqrt{s_{\rm NN}} = 130$ GeV
show for $p_\perp < 6$ GeV a suppression of 
neutral pion \cite{Adcox:2001jp} and charged hadron 
\cite{Adcox:2001jp,Adler:2002xw} transverse momentum spectra 
if compared to spectra in p+p collisions rescaled by the number
of binary collisions. This suppression is most pronounced (up to 
a factor $\sim 5$) in central Au+Au collisions and smoothly 
approaches the binary scaling case with decreasing centrality.
Within error bars, the suppression factors of $\pi^0$ and
charged hadron spectra agree, though central values for
the suppression of $\pi^0$ production are slightly 
lower \cite{Adcox:2001jp}. In addition,
a maximal azimuthal anisotropy $v_2(p_\perp)$ of hadroproduction
is found to persist up to the highest transverse 
momentum \cite{Adler:2002ct,Filimonov:2002xk,Ajitanand:2002qd}.
These data indicate the importance of final state medium effects 
up to $p_\perp < 6$ GeV.

Preliminary data shown at the Quark Matter 2002 conference confirm
these findings for Au+Au collisions at $\sqrt{s_{\rm NN}} = 200$ GeV;
they extend many observations up to $p_\perp \sim 10$ GeV.
In particular, data for the nuclear modification factor show an
approximately
constant maximal suppression within $6 < p_\perp < 12$ GeV for
charged hadrons \cite{Klay:2002xj,Jia:2002bk,Roland:2002un} and up
to $p_\perp < 8$ GeV for $\pi^0$ spectra 
\cite{d'Enterria:2002bw,Mioduszewski:2002wt}. The azimuthal anisotropy
$v_2(p_\perp)$ of charged hadrons stays close to maximal up to 
$p_\perp < 10$ GeV \cite{Kunde:2002pb}. Moreover, the disappearance
of back-to-back high-$p_\perp$ hadron 
correlations \cite{Adler:2002tq,Hardtke:2002ph,Chiu:2002ma}
provides an additional indication that final state medium effects
play a decisive role in hadroproduction up to $p_\perp \sim 10$ GeV.

{\it Theoretical situation:}
Parton energy loss 
has been proposed to account for the small nuclear modification 
factor \cite{Arleo:2002kh,Wang:2002ri,Jeon:2002dv}, the
azimuthal anisotropy 
\cite{Wang:2000fq,Gyulassy:2000gk,Lokhtin:2000wm,Vitev:2002pf}
and the disappearance of dijets \cite{Muller:2002fa,Hirano:2003hq}.
Quantitative studies indicate, however, that in the kinematic
regime relevant for RHIC ($p_\perp < 12$ GeV),  
$p_\perp$-broadening \cite{Hirano:2003hq,Wang:2002ri}, 
shadowing \cite{Jeon:2002dv,Wang:2002ri}, formation time \cite{Arleo:2002kh}
and possibly other effects contribute to the high-$p_\perp$
nuclear modification as well. Models have been proposed
which account for hadronic quenching without taking recourse to
parton energy loss. Instead, these models invoke 
string percolation \cite{Braun:2001us}, small hadronization
time arguments \cite{Gallmeister:2002us}, saturation
physics \cite{Kharzeev:2002pc}, the dominance of parton
recombination over parton fragmentation \cite{Fries:2003vb},
or initial state formation time arguments \cite{Lietava:2003df}.
The consistency and applicability of these models is currently
under debate. For the $p_\perp$-range accessible to RHIC, 
the competing hadronic effects may make it difficult to
disentangle {\it quantitatively} the contribution of parton energy
loss from the measured hadronic suppression pattern. The transverse 
phase space accessible to LHC ($E_\perp < 200$ GeV)
may turn out to be a qualitative advantage with this respect.

\begin{figure}[h]\epsfxsize=12.7cm
\centerline{\epsfbox{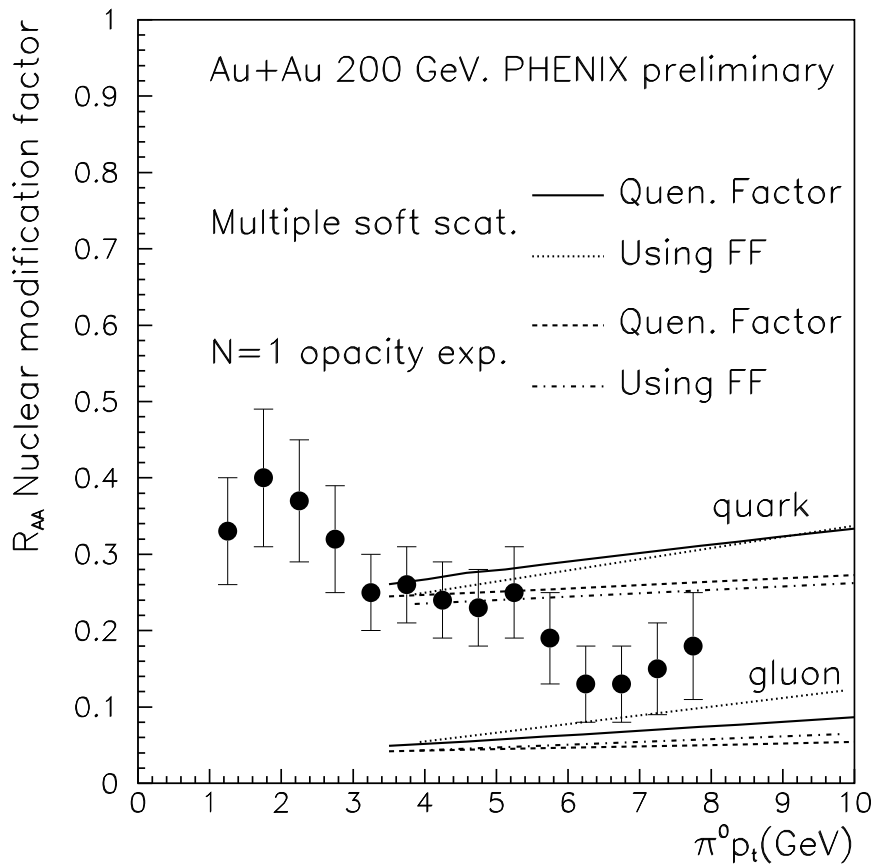}}
\vspace{0.5cm}
\caption{
The nuclear modification factor for $\pi^0$-production
\protect\cite{d'Enterria:2002bw,Mioduszewski:2002wt}
compared to model calculations involving parton energy loss
only. The lines show the quenching factor (\protect\ref{6.2})
with effective power $n=7$, and the suppression factor 
(\protect\ref{6.5}) calculated from medium-modified 
fragmentation functions. They are given in the
limiting cases where all parent partons are either quarks 
(upper lines) or gluons (lower lines). Calculations in
the multiple soft scattering approximation use $R = 2000$, 
$\omega_c = 67.5$ GeV, corresponding to $\hat{q} = 0.75
\frac{{\rm GeV}^2}{\rm fm}$ and $L=6$ fm. In the single
hard scattering approximation, we use $\bar{R} = R$, 
$\bar{\omega}_c = \omega_c$.
}\label{fig20}
\end{figure}

{\it Model comparison:}
In Fig.\ref{fig20}, we compare  
the two definitions (\ref{6.2}) and (\ref{6.5}) of the quenching
factor to the nuclear modification factor measured by the PHENIX
Collaboration in the 
$\pi^0$-spectra\cite{d'Enterria:2002bw,Mioduszewski:2002wt} of central
Au+Au collisions at $\sqrt{s_{\rm NN}} = 200$ GeV. We do not include
the nuclear modification factor for charged 
hadrons \cite{Klay:2002xj,Jia:2002bk,Roland:2002un} in Fig.\ref{fig20},
since charged hadrons are likely to be dominated at high $p_\perp$ by 
baryons whose production mechanism may involve additional
non-perturbative effects \cite{Vitev:2001zn}. 
\vspace{-0.5cm}
%
Our comparison makes
several simplifying assumptions which can be improved in further
studies: i) We do not model a realistic space-time geometry of the 
collision. Instead, we work for a fixed in-medium path length $L = 6$ fm. 
As a consequence, we do not compare to the centrality dependence 
of the nuclear modification factor for which a realistic distribution 
of in-medium path lengths and their impact parameter dependence is
needed. ii) We do not calculate the hard partonic matrix elements 
entering e.g. (\ref{6.3}). Instead, we determine the quenching 
factors directly from (\ref{6.2}) and (\ref{6.5}). As a consequence,
we do not know the $p_\perp$-dependent fractions of hard quarks and 
of hard gluons which fragment into a leading $\pi^0$. Since parton 
energy loss is different for quarks and gluons, we plot in 
Fig.\ref{fig20} the limits for which all parent partons are quarks
or gluons, respectively. The realistic curve lies in between
these limits. Since the ratio of parent quarks over parent
gluons increases with $p_\perp$, the $p_\perp$-dependence of the
realistic curve will be slightly steeper than the limiting cases
presented in Fig. \ref{fig20}.

From Fig.\ref{fig20}, we can draw several conclusions. First, the two
definitions of quenching factors in terms of hadronic spectra (\ref{6.2})
and in terms of fragmentation functions (\ref{6.5}) lead to 
quantitatively comparable results. Second, the multiple soft
and single hard scattering approximations for parton energy loss
lead to quantitatively comparable results. The slight variations
in $p_\perp$-slope should be regarded as theoretical uncertainties
in approximating (\ref{2.1}) and do not give any preference to
either approximation scheme.
Third, a calculation
based on partonic energy loss only can reproduce the magnitude
of the observed nuclear modification factor. Moreover, it results
in a very shallow $p_\perp$-dependence which seems consistent
with the current quality of experimental data. 
For an interpretation of the
model parameters used in Fig.\ref{fig20}, we use for a Bjorken 
scaling expansion $(\alpha = 1)$ the 
relation \cite{Salgado:2002cd,Gyulassy:2000gk} between the kinematic
constraint $R$ and the initially produced gluon density,
\begin{equation}
  R = \frac{L^2}{R_A^2}\, \frac{dN^g}{dy}\, ,
\label{6.7}
\end{equation}
where $R_A$ denotes the nuclear radius. The extracted value 
$\frac{dN^g}{dy} \simeq 2000$ is approximately a factor $2$
larger than a previous estimate \cite{Gyulassy:2000gk}
based on data from elliptic flow. Given the theoretical
uncertainties of parton energy loss calculations below 
$p_\perp < 10$ GeV, this factor $2$ mismatch constitutes
no inconsistency.

\section{Conclusion}
\label{sec7}

In recent years, phenomenological applications of medium-induced
parton energy loss were based mainly on two different approximations of
the medium-induced gluon energy distribution: the multiple soft
BDMPS-Z scattering approximation and the opacity approximation.
It remained unclear, however, to what extent these approximations
differ. Also, the quenching weights were studied mainly in kinematic
limiting cases.
Here, we have studied in both approximations the
medium-induced energy distributions (section \ref{sec2}), the
corresponding quenching weights (section \ref{sec3}), and the
extension of these calculations to dynamically expanding
collision region (section \ref{sec4}) and to the angular dependence
of the medium-induced radiation pattern (section \ref{sec5}).
The single hard scattering approximation is dominated by
the hard region $\omega > \bar{\omega}_c$ of
the gluon energy distribution while the multiple soft scattering
approximation is dominated by the soft region $\omega < \omega_c$.
Despite this difference, both approximations lead to
quantitatively comparable results if comparable sets of model
parameters are used. Numerically, we determine the 
correspondence
\begin{equation}
  \omega_c \simeq 3\, \bar{\omega}_c\, ,\qquad
  R \simeq 3\, \bar{R}\, ,
  \label{7.1}
\end{equation}
which relates the BDMPS transport coefficient $\hat{q}$ and the
Debye screening mass $\mu^2$ via an opacity $n_0\, L \sim 3$, 
see eq. (\ref{3.12}). Deviations from (\ref{7.1}) can be 
understood in terms of kinematic constraints on the
hard part of the gluon energy distribution (see discussion
of Fig. \ref{fig17}).

 The main result of this paper is the calculation of quenching
weights in section \ref{sec3}. We explained how to calculate
from these quenching weights the nuclear modification of hadronic
spectra. There are indications that the interpretation of RHIC data
on hadronic quenching requires additional physics effects beyond
the parton energy loss (see discussion in section \ref{sec6c}).
However, to discriminate energy loss contributions from these 
additional effects, a quantitatively reliable discussion of the 
current theoretical calculations of parton energy loss is needed. 
We hope that the CPU-inexpensive
subroutine for quenching weights which accompanies this paper
will prove a valuable tool to this end. Also, this routine can 
be used to explore observable consequences of parton energy loss 
in nucleus-nucleus collisions at LHC or for processes in cold nuclear matter.

{\bf Acknowledgment:} We thank Francois Arleo, Nestor Armesto,
Rolf Baier, Yuri Dokshitzer,
Kari Eskola, Alex Kovner, Andreas Morsch, J\"urgen Schukraft
and Boris Tomasik 
for helpful discussions. In particular, 
we thank Rolf Baier for suggesting the argument leading to eqs. 
(\ref{2.12}) and (\ref{2.13}). C.A.S. is supported by a Marie
Curie Fellowship of the European Community programme TMR
(Training and Mobility of Researchers), under the contract
number HPMF-CT-2000-01025.

%
\appendix

\section{The BDMPS-limit of the energy distribution (\ref{2.1})}
\label{appa}

Here, we establish that the $R\to \infty$ limit of the gluon energy
distribution (\ref{2.1}) coincides with the BDMPS result (\ref{2.9}). Using
the saddle point approximation (\ref{2.3}), the energy distribution
(\ref{2.1}) can be written in the form given in 
equations (A11) and (A12) of Ref.~\cite{Wiedemann:2000tf}. 
Integrated over transverse momentum $0< k_\perp < \chi\, \omega$,
one finds 
\begin{eqnarray}
\omega \frac{dI}{d\omega} &=& \frac{\alpha_s\, C_F}{\pi}
 \left( I_4 + I_5\right)\, ,
 \label{a1}\\ 
  I_4&=&{1\over 4 \omega^2}2 Re\int_0^Ldy_l\int_{y_l}^Ld\bar y_l\, 
  16A_4^2\nonumber \\
  && \times \left[\left(1-{iA_4B_4(\chi\omega)^2\over 4(D_4-iA_4B_4)^2}
            \right)\, \exp\left\{ -{(\chi\omega)^2
            \over 4(D_4-iA_4B_4)^2}\right\}-1\right] \, ,
  \label{a2}\\
  I_5 &=& {1\over\omega}Re\int_0^L dy_l{4A_5\over B_5}\left[\exp\left\{-
          {i(\chi\omega)^2\over 4A_5B_5}\right\}-1\right]\, ,
  \label{a3}
\end{eqnarray}
where
\begin{eqnarray}
  A_4  &=& \frac{\omega \Omega}{2\sin(\Omega (\bar{y}_L-y_L))}\, ,\qquad
  B_4 = \cos(\Omega (\bar{y}_L-y_L))\, ,
  \label{a4} \\
  D_4 &=& \frac{1}{2} n_0 C (L-\bar{y}_L)\, ,
  \label{a5}\\
  A_5 &=& \frac{\omega \Omega}{2\sin(\Omega y_L)}\, ,\qquad
  B_5 = \cos(\Omega y_L)\, ,
  \label{a6}
\end{eqnarray}
and 
\begin{equation}
 \Omega=(1+i)\, \sqrt{\frac{\hat{q}}{4\omega}}\, . 
  \label{a7}
\end{equation}
The limit $R\to\infty$
is obtained by taking $\chi\to\infty$ in eqs. (\ref{a2}) and (\ref{a3}):
\begin{eqnarray}
  \lim_{R\to\infty} I_4
   &=& -2Re\int_0^Ldy_l\int_{y_l}^Ld\bar y_l {\Omega^2\over
       \sin^2\left(\Omega\left(\bar y_l-y_l\right)\right)}
     \nonumber \\
   &=& - Re\int_0^Ldy_l\int_0^Ld\bar y_l {\Omega^2\over
       \sin^2\left(\Omega\left(\bar y_l-y_l\right)\right)}
     \nonumber \\
   &=& 2 Re\int_0^Ldy_l\, \Omega\, \frac{ \cos\left(\Omega y_l \right)}
       {\sin\left(\Omega y_l \right)}\, ,
   \label{a8}\\
  \lim_{R\to\infty} I_5
   &=& -2Re\int_0^L dy_l{\Omega\over \sin\left(\Omega y_l\right)
   \cos\left(\Omega y_l\right)}\, .
  \label{a9}
\end{eqnarray}
Both integrals are logarithmically divergent but this divergence
cancels in the sum
\begin{equation}
  \lim_{R\to\infty}\, 
  \left(I_4+I_5\right) = 2Re\, \ln\left[\cos\left(\Omega L\right)\right]\, .
  \label{a10}
\end{equation}
This coincides with the BDMPS result (\ref{2.9}).

\section{Gluon energy distribution to first order in opacity}
\label{appb}
In this appendix, we calculate the first order in opacity of the
gluon energy distribution (\ref{2.1}) for a Yukawa-type elastic
scattering center with Debye screening mass $\mu$:
\begin{equation}
  \vert a({\bf q})\vert^2 = \frac{\mu^2}{\pi ({\bf q}^2 + \mu^2)^2}\, .
  \label{b1}
\end{equation}
According to eq. (6.4) of Ref.\cite{Wiedemann:2000za}, the energy 
distribution takes the form:
\begin{eqnarray}
 \omega \frac{dI^{N=1}}{d\omega} &=& \frac{\alpha_s}{(2\pi)^2}
 \frac{2C_R}{\omega^2} \int_0^{\chi \omega} d{\bf k}\, 
 \int_0^\infty d{\bf q}\, \frac{\mu^2}{\pi ({\bf q}^2 + \mu^2)^2}\, 
 {\bf k}\cdot {\bf q}\, Z(Q,Q_1)\, ,
 \label{b2}
\end{eqnarray}
where
\begin{eqnarray}
 Q &=& \frac{{\bf k}^2}{2\omega}\, , \qquad
 Q_1 = \frac{({\bf k}+{\bf q})^2}{2\omega}\, ,
 \label{b3}
\end{eqnarray}
and
\begin{eqnarray}
 Z(Q,Q_1) &=& \lim_{\epsilon\to 0} {\rm Re}
 \int_{\xi_0}^\infty dy \int_{y}^\infty d\bar{y}\, 
 e^{-\epsilon y - \epsilon \bar{y}} 
 \nonumber \\
 && \qquad \times \int_y^{\bar{y}} d\xi\, n_0\, 
 \left( \frac{\xi_0}{\xi}\right)^\alpha\, 
 e^{-iQ(\bar{y}-\xi) - iQ_1(\xi - y)}
 \label{b4}\\
 &\equiv& \frac{n_0}{Q}\, \bar{Z}(Q_1)\, .
 \label{b5}
\end{eqnarray}
Irrespective of the value of the expansion parameter $\alpha$
in (\ref{b4}), the expression factorizes in the form (\ref{b5}).
In order to simplify (\ref{b2}), we shift the integration
variables by
 ${\bf q} \to {\bf q} - {\bf k}$, ${\bf k} \to {\bf k} 
\sqrt{\frac{2\omega}{L}}$ and ${\bf q} \to {\bf q} 
\sqrt{\frac{2\omega}{L}}$. This leads to 
\begin{eqnarray}
  \omega \frac{dI^{N=1}}{d\omega} &=& 4 \frac{\alpha_s\, C_R}{\pi}\, 
  \int_0^\kappa dk\, \int_0^{2\pi} d\varphi\, 
  \int_0^\infty q\, dq\, \frac{\partial}{\partial k} \left( 
  \frac{ 1}{k^2 +2kq\cos\varphi + q^2 + \gamma}\right)
  \nonumber \\
  && \qquad \times \left(\frac{\gamma}{2\pi}\right)\,
  \frac{n_0}{L}\,  \bar{Z}(q^2/L)
  \label{b6}
\end{eqnarray}
where
\begin{eqnarray}
  \gamma = \frac{\mu^2\, L}{2\, \omega}\, ,\qquad
  \kappa = \chi \sqrt{\frac{\omega L}{2}}
  = \sqrt{\frac{\bar R}{2\gamma}}\, .
  \label{b7}
\end{eqnarray} 
The $\varphi$- and $k$-integration in (\ref{b6}) can be done
analytically, 
\begin{eqnarray}
  \omega \frac{dI^{N=1}}{d\omega} &=& 4 \frac{\alpha_s\, C_R}{\pi}\,
   \frac{n_0}{L}\,   
  \int_0^\infty q\, dq\, \bar{Z}(q^2/L)
  \nonumber \\
  && \qquad \times
  \left( \frac{\gamma}{\sqrt{(\kappa^2 + q^2 + \gamma)^2
                       - 4 \kappa^2 q^2}}
         - \frac{\gamma}{q^2 + \gamma} \right)\, .
  \label{b8}
\end{eqnarray}
\begin{enumerate}
\item
\underline{For the static case, $\alpha = 0$}, the phase
factor (\ref{b5}) reads
\begin{equation}
  \bar{Z}_{\alpha=0}(Q_1) = \frac{-LQ_1 + sin(LQ_1)}{Q_1^2}\, ,
  \label{b9}
\end{equation}
and 
\begin{eqnarray}
  \omega \frac{dI_{\alpha=0}^{N=1}}{d\omega} &=& 4 \frac{\alpha_s\, C_R}{\pi}\,
   (n_0L)\, \gamma\,  
  \int_0^\infty q\, dq\, 
                  \frac{q^2 - sin(q^2)}{q^4}
                 \nonumber \\
     && \qquad \qquad \times
  \left( \frac{1}{q^2 + \gamma} - 
         \frac{1}{\sqrt{(\kappa^2 + q^2 + \gamma)^2
                       - 4 \kappa^2 q^2}}\right)\, .
  \label{b10}
\end{eqnarray}
Substituting $r = q^2$, we find (\ref{2.19}).
\item
\underline{For the Bjorken scaling case, $\alpha = 1$,} 
the phase (\ref{b5}) reads
\begin{eqnarray}
  && \hspace{-0.6cm} 
  \bar{Z}_{\alpha=1}(Q_1) = \frac{\xi_0}{Q_1}\, {\rm Re}\, 
  \left[ e^{iQ_1\xi_0}
         \left( {\rm Ei}[-iQ_1(L+\xi_0)] - {\rm Ei}[-iQ_1\xi_0] \right)
           + \ln \frac{\xi_0}{L+\xi_0} \right]
         \nonumber \\
         &&\hspace{1.2cm}= \frac{\xi_0}{Q_1}\, {\rm Re}\, 
         \left[ {\rm Ei}[-iQ_1L] 
                - \ln [-iQ_1L] - \gamma_E + O(\xi_0/L) \right]
                \nonumber \\
         &&\hspace{1.2cm}= \frac{\xi_0}{Q_1}\, {\rm Re}\, 
         \left[\int_0^{-iQ_1L}
         dt\, \frac{e^{-t}-1}{t} + O(\xi_0/L)\right] \, .
              \label{b11}
\end{eqnarray}
Here, $\gamma_E = 0.577 \dots$ is the
Euler constant and the exponential integral function ${\rm Ei}$ is 
defined in the text following eq. (\ref{4.3}).
Corrections of order $O(\xi_0/L)$ can be ignored since the time
of production $\xi_0$ is much smaller than the in-medium path length $L$.
With this approximation, one has
\begin{eqnarray}
  \omega \frac{dI_{\alpha=1}^{N=1}}{d\omega} 
   &=& 4 \frac{\alpha_s\, C_R}{\pi}\,
   (n_0\xi_0)\,   
  \int_0^\infty \frac{q\, dq}{q^2}\,
  {\rm Re}\left[- {\rm Ei}[-iq^2] + \ln [-iq^2] + \gamma_E \right]
 \nonumber \\
 && \qquad \times
  \left( \frac{\gamma}{q^2 + \gamma} - 
         \frac{\gamma}{\sqrt{(\kappa^2 + q^2 + \gamma)^2
                       - 4 \kappa^2 q^2}}\right)\, .
  \label{b12}
\end{eqnarray}
Substituting $r = q^2$, we find (\ref{4.3}).
\end{enumerate}

\section{The dipole approximation for an expanding medium}
\label{appc}
In this appendix, we follow Ref.~\cite{Baier:1998yf}
in giving explicit expressions for the path
integral (\ref{2.14}) in the dipole approximation,
\begin{eqnarray}
 &&{\cal K}({\bf r}_1,y_1;{\bf r}_2,y_2|\omega) =
 \int {\cal D}{\bf r}
   \exp\left[ i \frac{\omega}{2} \int_{y_1}^{y_2} d\xi
        \left(\dot{\bf r}^2
          - \frac{\Omega_\alpha^2(\xi_0)}{\xi^\alpha}\, {\bf r}^2 \right)
                      \right]\, .
  \label{c.1}
\end{eqnarray}
Equation (\ref{c.1})
is the path integral of
a 2-dimensional harmonic oscillator with time-dependent imaginary
frequency
\begin{equation}
  \frac{\Omega_\alpha^2(\xi_0)}{\xi^\alpha} 
  = \frac{\hat{q}(\xi)}{i\,2\,\omega}
  = -i \frac{\hat{q}_0}{2\omega} \left( \frac{\xi_0}{\xi}\right)^\alpha\, .
  \label{c.2}
\end{equation}
and {\it mass} $\omega$. The solution of (\ref{c.1}) can be written
in the form~\cite{Baier:1998yf}
\begin{eqnarray}
  {\cal K}({\bf r}_1,y_1;{\bf r}_2,y_2|\omega) =
  \frac{\omega}{2\pi\, i\, D(y_1,y_2)}\,
  \exp\left[ i S_{\rm cl}({\bf r}_1,y_1;{\bf r}_2,y_2) \right]\, .
  \label{c.3}
\end{eqnarray}
Here, the classical action $S_{\rm cl}$
in (\ref{c.3}) takes the form
\begin{eqnarray}
  S_{\rm cl}({\bf r}_1,y_1;{\bf r}_2,y_2)
  = \frac{\omega}{2} \left[ {\bf r}_{\rm cl}(\xi)
    \cdot \frac{d}{d\xi} {\bf r}_{\rm cl}(\xi)\right]
  \Bigg\vert^{y_1}_{y_2}\, ,
  \label{c.4}
\end{eqnarray}
where the classical path ${\bf r}_{\rm cl}(\xi)$ satisfies
the homogeneous differential equation
\begin{equation}
  \left[ \frac{ d^2}{d\xi^2}
          - \frac{\Omega_\alpha^2(\xi_0)}{\xi^\alpha}\right]\,
        {\bf r}_{\rm cl}(\xi)
        =0,
        \label{c.5}
\end{equation}
with initial conditions
\begin{equation}
  {\bf r}_{\rm cl}(y_1) = {\bf r}_1
  \, ,\hspace{20pt} \hbox{\rm and}\hspace{5pt}
  {\bf r}_{\rm cl}(y_2) = {\bf r}_2\, .
\label{c.6}
\end{equation}
The fluctuation determinant $D(\xi,\xi')$ in (\ref{c.3}) satisfies
\begin{equation}
  \left[ \frac{ d^2}{d\xi^2}
          - \frac{\Omega_\alpha^2(\xi_0)}{\xi^\alpha}\right]\,
        D(\xi,\xi')=0,
        \label{c.7}
\end{equation}
with initial conditions
\begin{equation}
  D(\xi, \xi)=0\, ,\hspace{20pt} \hbox{\rm and}\hspace{5pt}
  \frac{ d}{d\xi}   D(\xi,\xi')|_{\xi=\xi'}=1\, .
\label{c.8}
\end{equation}
In practice, $D(\xi,\xi')$ is found by combining the two independent
(scalar) solutions $f_1$, $f_2$ of (\ref{c.5}),
\begin{equation}
  D(\xi,\xi') = {\cal N}\, \left( f_1(\xi)\, f_2(\xi')
              - f_2(\xi)\, f_1(\xi') \right)\, ,
            \label{c.9}
\end{equation}
and fixing the norm ${\cal N}$ by the initial condition (\ref{c.8}).
The solution of (\ref{c.1}) can be written in terms of $D(\xi,\xi')$
and two $\xi$- and $\xi'$-dependent variables $c_1$, $c_2$,
\begin{eqnarray}
  {\cal K}({\bf r}_1,y_1;{\bf r}_2,y_2|\omega) &=&
  \frac{i\, \omega}{2\pi D(y_1,y_2)}\,
  \nonumber \\
  && \times \exp\left[-
    \frac{-i \omega}{2\, D(y_1,y_2)}
      \left(c_1 {\bf r}_1^2+ c_2 {\bf r}2^2 -
            2 {\bf r}_1\cdot {\bf r}_2\right)\right]\, .
  \label{c.10}
\end{eqnarray}
We consider three cases:
\begin{enumerate}
  \item \underline{The case $\alpha < 2$:}\\
    For this case, explicit expressions for (\ref{c.10}) are given
    in Appendix B of Ref.~\cite{Baier:1998yf}.
    The two independent solutions of the
    homogeneous differential equation (\ref{c.5}) are
    \begin{eqnarray}
      f_1(\xi) &=& \sqrt{\xi}\,
         I_\nu\left( 2\nu\, \Omega_\alpha(\xi_0)\,
             \xi^{\frac{1}{2\nu}}\right)\, ,
         \label{c.11} \\
      f_2(\xi) &=& \sqrt{\xi}\,
         K_\nu\left( 2\nu\, \Omega_\alpha(\xi_0)\,
                     \xi^{\frac{1}{2\nu}}\right)\, ,
         \label{c.12}
    \end{eqnarray}
    where $I_\nu$ and $K_\nu$ are modified Bessel functions
    with argument
\begin{equation}
  \nu = \frac{1}{2-\alpha} \, .
  \label{c.13}
\end{equation}
In terms of the variable [use $\Omega_\alpha(\xi_0) =
\sqrt{ -i \frac{\hat{q}_0}{2\,\omega}\xi_0^\alpha}$]
\begin{eqnarray}
  z(\xi) &=& 2 \nu\Omega_\alpha(\xi_0)\xi^{\frac{1}{2\nu}}\, ,
       \label{c.14}
\end{eqnarray}
the solution (\ref{c.10}) is given by ~\cite{Baier:1998yf}
[we use $z \equiv z(\xi)$,
$z' \equiv z(\xi')$]
\begin{eqnarray}
   D(\xi,\xi') &=&
   \frac{2\nu}{(2\nu\Omega_\alpha(\xi_0))^{2\nu}}\, (z z')^\nu
   \left[I_\nu(z)K_\nu(z')-K_\nu(z)I_\nu(z')\right]\, ,
   \label{c.15}\\
  c_1 &=& z\, \left(\frac{z'}{z}\right)^\nu\,
  \left[I_{\nu-1}(z)K_\nu(z')+K_{\nu-1}(z)I_\nu(z')\right]\, ,
  \label{c.16}\\
  c_2 &=& z'\, \left(\frac{z}{z'}\right)^\nu\,
  \left[K_\nu(z)I_{\nu-1}(z') + I_\nu(z)K_{\nu-1}(z')\right]\, .
  \label{c.17}
\end{eqnarray}
  \item \underline{The case $\alpha = 2$:}\\
    In this case, the two independent solutions of the
    homogeneous differential equation (\ref{c.5}) are
    $f_1(\xi) =  \xi^{\frac{1}{2}(1-A)}$ and
    $f_2(\xi) =  \xi^{\frac{1}{2}(1+A)}$
    where $A = \sqrt{1 + 4\Omega_{\alpha=2}^2(\xi_0)}$.
    From this, one finds
\begin{eqnarray}
   D(\xi,\xi') &=& \frac{1}{A}\, \left(\xi\, \xi'\right)^{\frac{1}{2}(1-A)}\,
   \left( \xi^A - \xi'^A\right)\, ,
   \label{c.18}\\
  c_1 &=& \frac{1+A}{A} \left(\frac{\xi}{\xi'}\right)^{-\frac{1}{2}(1-A)}
        - \frac{1-A}{A} \left(\frac{\xi'}{\xi}\right)^{\frac{1}{2}(1+A)}\, ,
  \label{c.19}\\
  c_2 &=& \frac{1+A}{A} \left(\frac{\xi'}{\xi}\right)^{-\frac{1}{2}(1-A)}
        - \frac{1-A}{A} \left(\frac{\xi}{\xi'}\right)^{\frac{1}{2}(1+A)}\, .
  \label{c.20}
\end{eqnarray}
\item \underline{The case $\alpha > 2$:}\\
In this case, the solution (\ref{c.15}) - (\ref{c.17}) has the
argument
\begin{eqnarray}
  z(\xi) &=& 2 \vert \nu\vert \Omega_\alpha(\xi_0)\xi^{\frac{1}{2\nu}}\, .
       \label{c.21}
\end{eqnarray}
Modified Bessel functions with negative index can be avoided with
the help of the identities: $K_{\nu}(z) = K_{-\nu}(z)$ and
$I_{\nu}(z) - I_{-\nu}(z) = - 2\frac{\sin(\nu \pi)}{\pi} K_{\nu}(z) $.

\begin{eqnarray}
   D(\xi,\xi') &=&
   2\nu \sqrt{\xi\, \xi'}
   \left[I_\nu(z)K_\nu(z')-K_\nu(z)I_\nu(z')\right]\, ,
   \label{c.22}\\
  c_1 &=& 2 \vert \nu\vert  \sqrt{\frac{\xi'}{\xi}}\,
  \xi^{\frac{1}{2\nu}} \, \Omega_\alpha(\xi_0)
  \left[I_{\nu-1}(z)K_\nu(z')+K_{\nu-1}(z)I_\nu(z')\right]\, ,
  \label{c.23}\\
  c_2 &=& 2 \vert \nu\vert  \sqrt{\frac{\xi}{\xi'}}\,
  \xi'^{\frac{1}{2\nu}} \, \Omega_\alpha(\xi_0)
  \left[K_\nu(z)I_{\nu-1}(z') + I_\nu(z)K_{\nu-1}(z')\right]\, .
  \label{c.24}
\end{eqnarray}
\end{enumerate}
With the solution (\ref{c.10}), the radiation spectrum can be written
as the sum of three contributions~\cite{Wiedemann:2000tf}
\begin{equation}
{dI^{\rm (tot)}\over d\omega}= \frac{1}{\omega}
 {d\sigma\over d\omega}
= {\alpha_s\over\pi^2} C_F (I_4+I_5+I_6)
= {dI^{\rm (vac)}\over d\omega} + {dI\over d\omega}\, .
\label{c.25}
\end{equation}
Here, $I_6=1/k_\perp^2$ is the medium-independent vacuum gluon
energy distribution. $I_4$ and $I_5$ determine the medium-induced
part ${dI\over d\omega}$ studied in this paper. They can be
computed in the dipole approximation:
\begin{eqnarray}
  I_4 &=& \frac{1}{4\omega^2}\,
  2{\rm Re} \int_{\xi_0}^{L+\xi_0} dy_l \int_{y_l}^{L+\xi_0} d\bar{y}_l
  \left( \frac{-4A^2_4 \bar{D}_4}{(\bar{D}_4-iA_4B_4)^2}
         + \frac{iA^3_4B_4\, {\bf k}_\perp^2}{(\bar{D}_4-iA_4B_4)^3} \right)
       \nonumber \\
       && \times
       \exp\left[{-\frac{{\bf k}_\perp^2}{4\, (\bar{D}_4 -
                  i\, A_4\, B_4)}}\right]\, ,
  \label{c.26}\\
  I_5 &=& \frac{1}{\omega}\, {\rm Re} \int_{\xi_0}^{L+\xi_0} dy_l\,
        \frac{-i}{B_5^2}\,
        \exp\left[{-i\frac{{\bf k}_\perp^2}{4\, A_5\, B_5}}\right]\, ,
          \label{c.27}
\end{eqnarray}
where
\begin{eqnarray}
    A_4&=&{\omega\over 2D(\bar y_l,y_l)}\, , \hspace{0.5cm}
    B_4=c_1(\bar y_l,y_l)\, , \hspace{0.5cm}
    \bar D_4 = \frac{1}{2}\int_{\bar y_l}^{L+\xi_0} d\xi
     n(\xi)\sigma({\bf r})
    \label{c.28}\\
    A_5&=&{\omega\over 2D(L+\xi_0,y_l)}\, , \hspace{0.5cm}
    B_5=c_1(L+\xi_0,y_l)\, .
    \label{c.29}
\end{eqnarray}
In the case $\alpha=0$, the functions
$I_{\pm 1/2}(z)$ and $K_{\pm 1/2}(z)$ entering (\ref{c.10})
have explicit expressions in terms of exponentials. One recovers
the known expressions for the static case
\cite{Zakharov:1996fv,Wiedemann:2000tf} from
\begin{eqnarray}
  A_4={\omega\Omega\over 2
      \sin(\Omega(\bar y_l-y_l))}\, , \hspace{1cm}
  B_4=\cos(\Omega(\bar y_l-y_l))\, , \hspace{1cm}
 \bar D_4=\frac{1}{2}n_0C(L-\bar y_l)\nonumber\, .
\end{eqnarray}


\end{document}